\documentclass[sigplan,10pt]{acmart}
\renewcommand\footnotetextcopyrightpermission[1]{} 
\usepackage[utf8]{inputenc}
\usepackage[american]{babel}

\usepackage{xspace}
\usepackage{multirow}
\usepackage{graphicx}
\usepackage{todonotes}
\usepackage{subfigure}
\usepackage{listings}
\usepackage{url}
\usepackage{balance}
\usepackage{amsfonts,amsmath,amsthm,amstext,latexsym}
\usepackage{paralist}
\usepackage[ruled,vlined,linesnumbered]{algorithm2e}
\usepackage{xcolor,colortbl}
\usepackage{makecell}
\usetikzlibrary{decorations.pathreplacing}

\newcommand{\ema}[1]{\ensuremath{#1}\xspace}
\newcommand{\VL}{\ema{\text{VL}}}

\newcommand{\ESC}{ESC\xspace}
\newcommand{\SPA}{SPA\xspace}
\newcommand{\SPARS}{SPARS\xspace}
\newcommand{\SPARSH}[1]{H-SPA(\ema{#1})\xspace}
\newcommand{\HASH}{HASH\xspace}
\newcommand{\HASHH}[1]{H-HASH(\ema{#1})\xspace}
\newcommand{\minb}{\ema{b_{min}}\xspace}
\newcommand{\maxb}{\ema{b_{max}}\xspace}

\newcommand{\EXPSPA}{\textsc{Spa}\xspace}
\newcommand{\EXPSPARS}{\textsc{Spars-40/40}\xspace}
\newcommand{\EXPSPARSVAR}{\textsc{Spars-16/64}\xspace}
\newcommand{\EXPSPARSH}{\textsc{H-Spa-40/40}\xspace}
\newcommand{\EXPSPARSHVAR}{\textsc{H-Spa-16/64}\xspace}
\newcommand{\EXPHASH}{\textsc{Hash-256/256}\xspace}
\newcommand{\EXPHASHVAR}{\textsc{Hash-32/256}\xspace}
\newcommand{\EXPHASHH}{\textsc{H-Hash-256/256}\xspace}
\newcommand{\EXPHASHHVAR}{\textsc{H-Hash-32/256}\xspace}
\newcommand{\EXPESC}{\textsc{ESC}\xspace}

\newcommand{\saylrt}{\texttt{saylr3}\xspace}
\newcommand{\saylrf}{\texttt{saylr4}\xspace}
\newcommand{\shermant}{\texttt{sherman3}\xspace}
\newcommand{\shermanf}{\texttt{sherman4}\xspace}
\newcommand{\watt}{\texttt{watt\_1}\xspace}
\newcommand{\gemat}{\texttt{gemat12}\xspace}
\newcommand{\lshp}{\texttt{lshp3466}\xspace}
\newcommand{\bcsp}{\texttt{bcspwr09}\xspace}
\newcommand{\dwt}{\texttt{dwt\_2680}\xspace}
\newcommand{\pores}{\texttt{pores\_2}\xspace}
\newcommand{\barth}{\texttt{barth4}\xspace}
\newcommand{\exdd}{\texttt{ex22}\xspace}
\newcommand{\lns}{\texttt{lns\_3937}\xspace}

\newcommand{\SPI}{\texttt{S40PI\_n1}\xspace}
\newcommand{\goodwin}{\texttt{Goodwin\_013}\xspace}
\newcommand{\oscil}{\texttt{oscil\_dcop\_30}\xspace}
\newcommand{\iprob}{\texttt{iprob}\xspace}
\newcommand{\nasa}{\texttt{nasa1824}\xspace}
\newcommand{\tols}{\texttt{tols1090}\xspace}
\newcommand{\str}{\texttt{str\_200}\xspace}
\newcommand{\bp}{\texttt{bp\_0}\xspace}
\newcommand{\olm}{\texttt{olm1000}\xspace}
\newcommand{\tub}{\texttt{tub1000}\xspace}
\newcommand{\young}{\texttt{young3c}\xspace}
\newcommand{\fpga}{\texttt{fpga\_dcop\_05}\xspace}
\newcommand{\qh}{\texttt{qh1484}\xspace}
\newcommand{\rdb}{\texttt{rdb1250}\xspace}
\newcommand{\adder}{\texttt{adder\_dcop\_01}\xspace}
\newcommand{\poli}{\texttt{poli}\xspace}
\newcommand{\koho}{\texttt{Kohonen}\xspace}
\newcommand{\ors}{\texttt{orsreg\_1}\xspace}
\newcommand{\wang}{\texttt{wang1}\xspace}
\newcommand{\shyy}{\texttt{shyy41}\xspace}
\newcommand{\rw}{\texttt{rw5151}\xspace}
\newcommand{\hamr}{\texttt{Hamrle2}\xspace}
\newcommand{\gresley}{\texttt{LeGresley\_4908}\xspace}
\newcommand{\rajat}{\texttt{rajat03}\xspace}
\newcommand{\cage}{\texttt{cage9}\xspace}
\newcommand{\dw}{\texttt{dw1024}\xspace}
\newcommand{\cheby}{\texttt{Chebyshev3}\xspace}

\definecolor{red}{RGB}{255,51,51}
\definecolor{SPARS}{RGB}{255,255,255}
\definecolor{SPA}{RGB}{255,255,255}
\definecolor{best}{gray}{0.90}
\definecolor{sbest}{gray}{0.90}
\definecolor{worst}{gray}{0.65}

\newcommand{\val}[1]{{\color{blue}#1}}

\newcommand{\shepherd}[1]{#1}

\graphicspath{{figs/}}

\hypersetup{draft}
\begin{document}
 
\title{Optimization of SpGEMM with Risc-V vector instructions}

\author{Valentin Le Fèvre}
\affiliation{\institution{Barcelona Supercomputing Center}
\city{Barcelona}
\country{Spain}}
\email{valentin.lefevre@bsc.es}
\author{Marc Casas}
\affiliation{\institution{Barcelona Supercomputing Center}
\city{Barcelona}
\country{Spain}}
\email{marc.casas@bsc.es}


\setcopyright{none}

\begin{abstract}
The Sparse GEneral Matrix-Matrix multiplication (SpGEMM) $C=A \times B$ is a fundamental routine 
extensively used in domains like machine learning or graph analytics.
Despite its relevance, the efficient execution of SpGEMM on vector architectures is a relatively unexplored topic.
The most recent algorithm to run SpGEMM on these architectures is based on the SParse Accumulator (SPA) approach, and it is relatively efficient for sparse matrices featuring several tens of non-zero coefficients per column as it computes $C$ columns one by one.
However, when dealing with matrices containing just a few non-zero coefficients per column, the state-of-the-art algorithm is not able to fully exploit long vector architectures when computing the SpGEMM kernel.

To overcome this issue we propose the SPA paRallel with Sorting (\SPARS) algorithm, which computes in parallel several $C$ columns among other optimizations, and the \HASH algorithm, which uses dynamically sized hash tables to store 
intermediate output values.
To combine the efficiency of \SPA for relatively dense matrix blocks with the high performance that \SPARS and \HASH deliver for very sparse matrix blocks we propose \SPARSH{t} and \HASHH{t}, which dynamically switch between different algorithms.
\SPARSH{t} and \HASHH{t} obtain 1.24$\times$ and 1.57$\times$ average speed-ups with respect to \SPA respectively, over a set of 40 sparse matrices obtained from the SuiteSparse Matrix Collection~\cite{suitesparseMC}.
For the 22 most sparse matrices, \SPARSH{t} and \HASHH{t} deliver 1.42$\times$ and 1.99$\times$ average speed-ups respectively.

\end{abstract}

\keywords{vector processor, sparse matrix, sparse multiplication, Risc-V, SpGEMM}

\maketitle
\pagestyle{plain} 

 \section{Introduction}
 \label{sec.intro}
 SpGEMM (Sparse GEneral Matrix-Matrix multiplication) is one of the most fundamental routines in sparse linear algebra.
It is widely used in sparse linear solvers~\cite{multigrid}, graph analytics~\cite{triangle_counting} or machine learning~\cite{sparse_dnn}.
This routine uses two sparse matrices $A$ and $B$ 
to compute $C=A\times B$.
Since both input matrices are sparse, SpGEMM displays much more irregular memory access patterns than the  
Sparse Matrix Multi-vector multiplication (SpMM) or the Sparse Matrix Dense Matrix multiplication (SpMDM), where $B$ is a dense matrix.
The efficient execution of SpGEMM on many-core architectures~\cite{spgemm_manycore,spgemm_manycore2}, GPUs~\cite{dense_gpu,hash_example,spgemm_tiled},
or both~\cite{hash_example2}, has been extensively studied.

Hardware designs supporting vector Instruction Set Architectures (ISAs), like the ARM Scalable Vector Extension (SVE)~\cite{stephens2017arm} or the RISC-V "V" vector extensions~\cite{riscvv}, are becoming increasingly attractive~\cite{Rico17}.
Commercial products featuring long vector processors like the Aurora-SX processor~\cite{yamada2018vector} confirm this trend towards using increasingly long vector units, 
which are able to manipulate multiple elements within a single instruction, for high-end computing.
The number of elements managed by a single vector instruction is called the{\it Vector Length} (\VL). \shepherd{\VL can be dynamically set 
to any value between 1 and the maximum number of elements that registers can store.}

Despite the relevance of vector processors for high-end computing, the most recent SpGEMM algorithm targeting these architectures~\cite{spgemm_vector}, which is based on the {\it SParse
Accumulator} (\SPA) approach~\cite{dense_matlab}, presents remarkable limitations. 
While this algorithm is relatively efficient when manipulating matrices that feature tens of non-zero coefficients per column,
it is unable to fully exploit long vector architectures when dealing with very sparse matrices column as it computes $C$ columns one by one.

To overcome this limitation, we propose two SpGEMM algorithms targeting long vector architectures: 
First, the {\it \SPA paRallel with Sorting (\SPARS)} algorithm, which is able to 
exploit the potential of large vector lengths when manipulating very sparse matrices 
and uses a dense array to keep the intermediate values of the output matrix. The main novelty of \SPARS with respect to previous work~\cite{spgemm_vector} is the use of a brand new {\it blocking} algorithm based on a sorting process. 
Second, we propose the {\it \HASH} algorithm, which uses hash tables to store intermediate values of the output matrix instead of a large dense array.
\HASH leverages the same blocking mechanism as \SPARS and it also uses a novel approach to dynamically adapt the size of its hash table.
In addition, to combine the efficiency of \SPARS and \HASH for very sparse matrices and the good performance of state-of-the-art \SPA-based algorithms when manipulating matrices with tens of non-zero coefficients per column, we propose two new algorithms: \SPARSH{t} and the \HASHH{t}.
They exploit a hybrid execution model, $i.e.$, \SPARSH{t} dynamically switches between \SPA and \SPARS while \HASHH{t} does so between \SPA and \HASH.
Overall, this paper makes the following contributions over the state-of-the-art:

\begin{itemize}
\item We propose the \SPARS and \HASH algorithms, which are able to efficiently compute the SpGEMM operation 
on long vector architectures for very sparse matrices.
\SPARS uses a large data structure to keep the intermediate values of the output matrix, while \HASH employs a more sophisticated approach based on hash tables.
\item We propose the hybrid \SPARSH{t} and \HASHH{t} algorithms, which dynamically combine \SPARS and \HASH, respectively, with \SPA. This combination is driven by the sparsity structure of each matrix block. 
\item We evaluate \SPA, \SPARS, and \SPARSH{t}, \HASH, \HASHH{t}, and an algorithm based on sorting (\ESC~\cite{esc_gpu_dalton,winter_ESC}) considering a vector processor architecture 
featuring eight vector lanes and supporting a maximum \VL of 256 double-precision elements. 
\SPARSH{t} and \HASHH{t} obtain 1.24$\times$ and 1.57$\times$ average speed-ups with respect to \SPA, respectively, over a set of 40 sparse matrices of the SuiteSparse Matrix Collection~\cite{suitesparseMC}.
For the 22 most sparse matrices, \SPARSH{t} and \HASHH{t} deliver 1.42$\times$ and 1.99$\times$ average speed-ups respectively.
\end{itemize}


 \section{Approaches to run SpGEMM on vector architectures}
 \label{sec.algo}
 This section describes some previously proposed approaches for the SpGEMM operation.
Section~\ref{sec.algo.intro} describes the Gustavson method~\cite{gustavson}.
Section~\ref{sec.algo.SPA} explains the SParse Accumulator (SPA) approach~\cite{dense_matlab} and its corresponding vectorization, which is based on previous work~\cite{spgemm_vector}.



\vspace{-0.2cm}
\subsection{Gustavson method}
\label{sec.algo.intro}

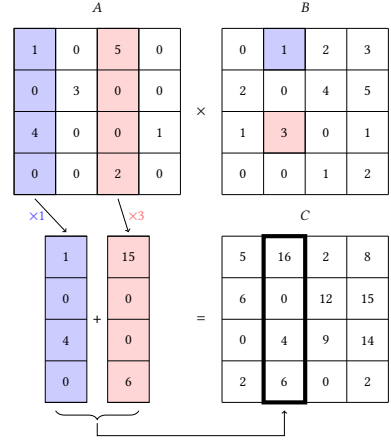
\begin{figure}
 
\resizebox{0.6\linewidth}{!}{
\begin{tikzpicture}
    \draw[fill=blue!20!white] (6,4) rectangle (7,3);
    \draw[fill=red!20!white] (6,2) rectangle (7,1);
    
    \draw[fill=blue!20!white] (0,4) rectangle (1,0);
    \draw[fill=red!20!white] (2,4) rectangle (3,0);
    
    \draw[fill=blue!20!white] (0.75,-1) rectangle (1.75,-5);
    \draw[fill=red!20!white] (2.25,-1) rectangle (3.25,-5);
    
    \draw[line width=1mm] (6,-1) rectangle (7,-5);
 \foreach \x in {0,...,4}
 {
    \draw (0,\x) -- (4,\x);
    \draw (\x,0) -- (\x,4);
    
    \pgfmathtruncatemacro{\y}{\x + 5};
    \draw (5,\x) -- (9,\x);
    \draw (\y,0) -- (\y,4);

    \pgfmathtruncatemacro{\y}{\x + 5};
    \pgfmathtruncatemacro{\z}{\x - 5};
    \draw (5,\z) -- (9,\z);
    \draw (\y, -5) -- (\y, -1);
    
    \draw (0.75,\z) -- (1.75,\z);
    \draw (2.25,\z) -- (3.25,\z);
    
  }
  
    \node at (2,4.5) {$A$};
    \node at (7,4.5) {$B$};
    \node at (7,-0.5) {$C$};
    \node at (4.5,2) {$\times$};
    \node at (2,-3) {$+$};
    \node at (4.5,-3) {$=$};
    \node at (0.5,0.5) {0};
    \node at (1.5,0.5) {0};
    \node at (2.5,0.5) {2};
    \node at (3.5,0.5) {0};
    \node at (0.5,1.5) {4};
    \node at (1.5,1.5) {0};
    \node at (2.5,1.5) {0};
    \node at (3.5,1.5) {1};
    \node at (0.5,2.5) {0};
    \node at (1.5,2.5) {3};
    \node at (2.5,2.5) {0};
    \node at (3.5,2.5) {0};
    \node at (0.5,3.5) {1};
    \node at (1.5,3.5) {0};
    \node at (2.5,3.5) {5};
    \node at (3.5,3.5) {0};
    \node at (5.5,0.5) {0};
    \node at (6.5,0.5) {0};
    \node at (7.5,0.5) {1};
    \node at (8.5,0.5) {2};
    \node at (5.5,1.5) {1};
    \node at (6.5,1.5) {3};
    \node at (7.5,1.5) {0};
    \node at (8.5,1.5) {1};
    \node at (5.5,2.5) {2};
    \node at (6.5,2.5) {0};
    \node at (7.5,2.5) {4};
    \node at (8.5,2.5) {5};
    \node at (5.5,3.5) {0};
    \node at (6.5,3.5) {1};
    \node at (7.5,3.5) {2};
    \node at (8.5,3.5) {3};
    
    \node at (1.25,-1.5) {1};
    \node at (1.25,-2.5) {0};
    \node at (1.25,-3.5) {4};
    \node at (1.25,-4.5) {0};
    \node at (2.75,-1.5) {15};
    \node at (2.75,-2.5) {0};
    \node at (2.75,-3.5) {0};
    \node at (2.75,-4.5) {6};

    \node at (5.5,-4.5) {2};
    \node at (6.5,-4.5) {6};
    \node at (7.5,-4.5) {0};
    \node at (8.5,-4.5) {2};
    \node at (5.5,-3.5) {0};
    \node at (6.5,-3.5) {4};
    \node at (7.5,-3.5) {9};
    \node at (8.5,-3.5) {14};
    \node at (5.5,-2.5) {6};
    \node at (6.5,-2.5) {0};
    \node at (7.5,-2.5) {12};
    \node at (8.5,-2.5) {15};
    \node at (5.5,-1.5) {5};
    \node at (6.5,-1.5) {16};
    \node at (7.5,-1.5) {2};
    \node at (8.5,-1.5) {8};
 
    \draw[decorate,decoration={brace,amplitude=6pt}] (3,-5.2) -- (1,-5.2); 
    \draw[->] (2,-5.5) -- +(0,-0.3) -| (6.5,-5.2);
    \draw[->] (0.5,-0.1) -- (1.25,-0.9) node[midway,left,text=blue!70!white] {$\times 1$};
    \draw[->] (2.5,-0.1) -- (2.75,-0.9) node[midway,right,text=red!70!white] {$\times 3$};
 
\end{tikzpicture}
}

\caption{Gustavson's algorithm for the second column.\label{fig.gustavson}}
\end{figure}

The Gustavson method~\cite{gustavson} for columns
multiplies two sparse matrices $A$ and $B$, that is, computes $C=A\times B$, where $A$ is a matrix of size $m_A\times n_A$ and $B$ is a matrix of size $n_A\times n_B$.
\shepherd{We represent these matrices using the Compressed Sparse Column (CSC) format~\cite{CSC_ref}.
For example, for matrix $B$ with $NNZ$ non-zero elements, the CSC format stores the matrix using 3 arrays:
\begin{itemize}
\item one array \textit{values} of size $NNZ$ that stores the value of each non-zero element;
\item one array \textit{row\_indices} of size $NNZ$ that stores the corresponding row index of each non-zero element;.
\item one array \textit{column\_pointers} of size $n_B+1$ containing indices that correspond to the beginning of each column in the two previous arrays. The first cell always contains 0, and the last cell always contains $NNZ$.
\end{itemize}
 With the CSC format, the order in which elements are stored in \textit{values} and \textit{row\_indices} is not important in the same column, but the columns have to be stored in the order so that
 \textit{column\_pointers}[i+1]-\textit{colum\_pointers}[i] always corresponds to the number of non-zero elements in column i.
}
This method considers each $C$ column 
as a linear combination of $A$ columns. 
More formally, to compute column $j$ of $C$, the algorithm multiplies each non-zero element $B_{kj}$ by column $k$ of $A$ and
accumulate these values. 
Figure~\ref{fig.gustavson} provides an example of the Gustavson method considering $4\times4$ matrices.
This example computes the second column of $C$ by i) multiplying the first of column of $A$ by the $B_{01}$ coefficient, which has a numerical value of 1, and the third column of $A$ by the $B_{03}$ coefficient; and ii) obtaining the final $C$ column by adding the two columns obtained in the first step.
This is a highly parallel method since each $C$ column can be computed independently of the others.
Its main problem is load imbalance, as computations
of two different columns do not necessarily have the same number of operations to perform.
\shepherd{Note that the same method could be applied on rows instead of columns, with matrices represented using the Compressed Sparse Row (CSR) format.}

There are several variations of the Gustavson method that compute the final $C$ column following different approaches.
When an element $A_{ik} \times B_{kj}$ is computed, one needs to know whether it results in a new non-zero element in column $j$ of $C$
or it needs to be accumulated with already computed values $A_{il}\times B_{lj}$ for any $l \neq k$.
This operation can be carried out by a hash table~\cite{hash_example,hash_example2}, by sorting and merging keys~\cite{merge_dalton,sort_bell}, or by the use of a dense vector, that is, a dense data structure that stores all intermediate accumulated values~\cite{dense_gpu,dense_matlab}. 
Section~\ref{sec:proposal} proposes two new algorithms based on the dense vector and the hash table accumulators, respectively, to efficiently run SpGEMM on vector processors. 

%

 \subsection{SParse Accumulator (SPA)}
 \label{sec.algo.SPA}
 
\setlength{\textfloatsep}{6pt}

\begin{algorithm}
  \SetKwInOut{Input}{Input}
  \SetKwData{SPAV}{\texttt{SPA\_values}}
  \SetKwData{SPAF}{\texttt{SPA\_flags}}
  \SetKwData{SPAI}{\texttt{SPA\_indices}}
  \SetKwData{cpt}{\texttt{counter}}
  \SetKwData{val}{\texttt{tmp\_val}}
  \SetKwFunction{Append}{append}
  \Input{Index $j$ of the column}
  Init \SPAV, \SPAF, \SPAI with zeros\;
  $\cpt \leftarrow 0$\;
  \ForEach{element $B_{kj}$ in column $j$ of $B$}
  {
    \ForEach{element $A_{ik}$ in column $k$ of $A$}
    {
      $\val \leftarrow A_{ik}\times B_{kj}$\;
      $\SPAV[i]$ += \val\;
      \If{$\SPAF[i]$ == 0}{
         $\SPAF[i] \leftarrow 1$\;
         $\SPAI[\cpt] \leftarrow i$\;
         Increment \cpt\;
      }
    }
  }
  \For{$i \leftarrow 0$ \KwTo $\cpt-1$}
  {
    \Append{C.\texttt{values}, $\SPAV[\SPAI[\cpt]]$}\;
    \Append{C.\texttt{row\_indices}, $\SPAI[\cpt]$}\;
  }
  C.\texttt{column\_pointers[j+1]} $\leftarrow$ C.\texttt{column\_pointers[j]}+\cpt\;
  \caption{Sequential \SPA algorithm}\label{algo.spa}
 \end{algorithm}

 This section describes the SParse Accumulator algorithm (\SPA)~\cite{dense_matlab}. 
 The sequential SPA algorithm to compute $C=A\times B$ is described in Algorithm~\ref{algo.spa}. For simplicity, we show a pseudo-code that computes the $j$-th column of the output $C$ matrix, although this pseudo-code can be trivially generalized to any generic number of columns.
 The first step is to create three arrays with the same size as the C row count:
 \texttt{SPA\_values}, \texttt{SPA\_flags} and \texttt{SPA\_indices}. 
 The \texttt{SPA\_values} array stores partial values of C non-zero coefficients, 
 \texttt{SPA\_flags} describes whether or not SPA has already computed a non-zero contribution for a certain coefficient of the $j$-th C column, 
 and \texttt{SPA\_indices} stores indices of C elements for which a non-zero contribution has already been obtained.
 SPA initializes these three arrays and a counter of non-zero elements with zeros (lines 1 and 2 of Algorithm~\ref{algo.spa}).

 SPA iterates over all non-zero elements in column $j$ of $B$ and multiplies each non-zero $B_{kj}$ coefficient by all its corresponding non-zero values $A_{ik}$ present in column $k$ of $A$. 
 After multiplying non-zero elements $A_{ik}$ and $B_{kj}$ (line 5 of Algorithm~\ref{algo.spa}),
 the SPA algorithm adds the value $A_{ik}\times B_{kj}$ to the $i$-th position of \texttt{SPA\_values} (line 6 of Algorithm~\ref{algo.spa}). 
 Then, if \texttt{SPA\_flags[i]} is zero, SPA changes this value to 1 to indicate that a non-zero contribution to the $i$-th position of the $j$-th C column exists.
 Also, SPA stores row index $i$ to position \texttt{counter} in \texttt{SPA\_indices}, and increments the \texttt{counter} value to indicate that a new non-zero output coefficient has been created (lines 9 and 10 of Algorithm~\ref{algo.spa}).
 Otherwise, when the \texttt{SPA\_flags[i]} entry value is one, SPA does not update array \texttt{SPA\_indices} and variable \texttt{counter} since the corresponding $i$-th position of the $j$-th C column already has a non-zero contribution. 
 Finally, once SPA finishes all iterations over the non-zero elements of the $B$ $j$-th column, the algorithm updates the non-zero values corresponding to column $j$ of C.
 Variable \texttt{counter} contains the total number of non-zero elements to be added to the C.\texttt{values} and C.\texttt{row\_indices} data structures. 


 
 We vectorize \SPA applying the most recent approaches to run SpGEMM on long vector architectures~\cite{spgemm_vector}. 
 Write dependencies are a key factor to obtain an efficient vectorization.
 Write dependencies only occur when there are two
 indices $k$ and $l$ such that $A_{ik}B_{kj}$ and $A_{il}B_{lj}$ are not equal to 0. 
 However, for one non-zero $B_{kj}$, all multiplications
 by elements of column $k$ of $A$ have different row indices, resulting in totally independent write operations in the dense vectors.
 Therefore, it is possible to vectorize SPA as follows: we iterate over all non-zero elements of a column of $B$ and for each element, using vector instructions, we load an entire column of $A$.
 The current values of \texttt{SPA\_values} and \texttt{SPA\_flags} are retrieved, then the column of $A$ is multiplied by the current non-zero in $B$, accumulated and stored back. Finally, a test
 of the flags is performed and new indices are stored contiguously in \texttt{SPA\_indices}. Algorithm~\ref{algo.spa_vec} represents this SPA vectorization. All variables whose name starts
 with \texttt{v} are stored in vector registers. 
We describe each vector instruction at the right-hand side of each line.

 
  \begin{algorithm}
  \SetKwInOut{Input}{Input}
  \SetKwData{SPAV}{\texttt{SPA\_values}}
  \SetKwData{SPAF}{\texttt{SPA\_flags}}
  \SetKwData{SPAI}{\texttt{SPA\_indices}}
  \SetKwData{cpt}{\texttt{counter}}
  \SetKwFunction{Append}{append}
  \Input{Index $j$ of the column}
  Init \SPAV, \SPAF, \SPAI with zeros\;
  \ForEach{element $B_{kj}$ in column $j$ of $B$}
  {
    $\VL \leftarrow $ A.\texttt{column\_pointers[k+1]}-A.\texttt{column\_pointers[k]}\tcp*[r]{Set Vector Length} 
    vA $\leftarrow$ Load column $k$ of $A$\tcp*[r]{Contiguous Vector Load}
    vSPA $\leftarrow$ Load corresponding elements of \SPAV\tcp*[r]{Indexed Vector Load}
    vSPA\_flags $\leftarrow$ Load corresponding elements of \SPAF\tcp*[r]{Indexed Vector Load}
    vSPA $\leftarrow$ vSPA + vA $\cdot$~$B_{kj}$\tcp*[r]{Vector Fused Multiply-Add}
    Store back vSPA\tcp*[r]{Indexed Vector Store}
    Compare vSPA\_flags to 0 to create a mask\tcp*[r]{Create Vector Mask}
    Store vSPA\_indices using the mask\tcp*[r]{Indexed Vector Store}
  }
  Store the column as sparse\;
  \caption{Vectorized \SPA algorithm}\label{algo.spa_vec}
 \end{algorithm}

There are two important considerations to make regarding Algorithm~\ref{algo.spa_vec}:
First, while Algorithm~\ref{algo.spa_vec} assumes that the maximum VL is larger than the maximum number of non-zeros per $A$ column (Line 3), it is trivial to modify it to handle the general case where a column of $A$ may have more non-zero elements than the maximum available \VL.
%
 To support this scenario the algorithm just needs to iterate several times over the same column, which only requires adapting the indices of the $A$ elements to load (line 4 of Algorithm~\ref{algo.spa_vec}).
Second,
 a masked vectorized store, which could be used at line 11 to store non-zero values and indices, does not store the unmasked elements contiguously.
Instead, there are several alternatives to store unmasked non-zero elements like using vector instructions to store in a vector register just the unmasked elements and then perform a regular vector store, or computing the indices of the unmasked elements and apply an index stored instruction on them.

The vectorized \SPA algorithm only exploits SIMD parallelism across $A$ columns since it processes $B$ non-zero elements one by one.
Since sparse matrices typically feature a low number of non-zero elements per column, the vectorized \SPA algorithm may not able to fully exploit vector processors featuring long vector lengths.
Section~\ref{sec.expe} shows experimental results that confirm this limitation of \SPA.
A new algorithm able to simultaneously process several $A$ and $B$ non-zero elements is required to fully exploit long vector architectures.

 \section{Proposals to run SpGEMM with very sparse matrices on vector architectures}
 \label{sec:proposal}
 \subsection{SPA paRallel with Sorting (SPARS)}
 \label{sec.algo.SPARS}
 The SPA paRallel with Sorting (SPARS) algorithm computes the SpGEMM kernel $C=A \times B$ and it is able to fully exploit long vector architectures even for largely sparse matrices featuring a few non-zero elements per column.
 SPARS exploits that computations involving different $B$ columns do not have dependencies, which is an aspect of the SpGEMM kernel that has been already exploited in the context of distributed memory architectures~\cite{spgemm_distributed}.
 \shepherd{By handling one column per vector register element, \SPARS is able to simultaneously process as many columns as the maximum \VL.
 \SPARS splits matrix $B$ into several blocks and processes them sequentially. 
 The number of columns per block drives the vector length.
 This section describes a \emph{Blocking Algorithm} to determine the size of each block.}
 To avoid write dependencies, each $B$ column has its own dense vectors, that is, \texttt{SPA\_values}, \texttt{SPA\_flags}, and \texttt{SPA\_indices}
 are now dense matrices of size $m_A \times \VL$, where \VL represents the vector length and $m_A$ is the row count of $A$ and $C$.

 Algorithm~\ref{algo.spars} shows the pseudo-code of \SPARS~\shepherd{to process one block of columns}.
 It computes \VL $C$ columns starting at column $j$. 
 For some vector instructions we use the $i$ index to specify what is computed for each scalar element.
 Thus, each line referencing $i$ should be understood
 as an implicit \textit{for all $i$ from 0 to $\VL-1$}.
 \SPARS loads values from $B$ and $A$ matrices (lines 7 and 8 of Algorithm~\ref{algo.spars}) to multiply and accumulate them in registers storing the \texttt{SPA\_values} data (line 11), similarly as \SPA does.
 \SPARS uses a vector register (\texttt{vIndices\_B}) to store indices of $B$ columns non-zeros.
 \SPARS initializes \texttt{vIndices\_B} with the indices of 
 the first element in each column, and determines whether it has reached the end of a $B$ column when the indices reach the beginning of the next column, which has its indices stored in the \texttt{vEnd\_B} register. 
 \SPARS also keeps track of positions in the $A$ columns in the
 \texttt{vCounter\_A} register, which is reset when \SPARS reaches the end of the corresponding $A$ column (line 16). 
 In this case, \SPARS moves to the next non-zero element in the $B$ column by increasing the values in \texttt{vIndices\_B}.
 In the end, \SPARS stores the columns using the same vectorized algorithm as \SPA.
 
 \begin{algorithm}
  \SetKwInOut{Input}{Input}
  \SetKwData{SPAV}{\texttt{SPA\_values}}
  \SetKwData{SPAF}{\texttt{SPA\_flags}}
  \SetKwData{SPAI}{\texttt{SPA\_indices}}
  \Input{Index $j$ of first column and \VL, number of columns to process}
  Init \SPAV, \SPAF, \SPAI with zeros\;
  vIndices\_B[i] $\leftarrow$ B.\texttt{column\_pointers[j+i]}\tcp*[r]{Contiguous Vector Load}
  vEnd\_B[i] $\leftarrow$ B.\texttt{column\_pointers[j+i+1]}\tcp*[r]{Contiguous Vector Load}
  vCounter\_A[i] $\leftarrow$ 0\tcp*[r]{Vector Broadcast}
  Create a mask vMask to keep elements where vIndices\_B[i] $<$ vEnd\_B[i]\tcp*[r]{Create Vector Mask}
  \While{there is one unmasked element in vMask}{
    vB $\leftarrow$ Load elements of $B$ at index vIndices\_B\tcp*[r]{Indexed Vector Load}
    vA[i] $\leftarrow$ Load vCounter\_A[i]-th element of corresponding column in $A$\tcp*[r]{Indexed Vector Load}
    vSPA $\leftarrow$ Load corresponding elements of \SPAV\tcp*[r]{Indexed Vector Load}
    vSPA\_flags $\leftarrow$ Load corresponding elements of \SPAF\tcp*[r]{Indexed Vector Load}
    vSPA $\leftarrow$ vSPA + vA $\cdot$ vB\tcp*[r]{Vector Fused Multiply-Add} 
    Store back vSPA\tcp*[r]{Indexed Vector Store}
    Compare vSPA\_flags to 0 to create a mask vFlags\tcp*[r]{Create Vector Mask}
    Store vSPA\_indices using vFlags\tcp*[r]{Indexed Vector Store}
    \eIf{vA[i] was the last element of the column}{
      vCounter\_A[i] $\leftarrow$ 0\tcp*[r]{Vector Broadcast}
      vIndices\_B[i] $\leftarrow$ vIndices\_B[i]+1\tcp*[r]{Vector Add}
    }{
      vCounter\_A[i] $\leftarrow$ vCounter\_A[i]+1\tcp*[r]{Vector Add}
    }
    Update vMask\tcp*[r]{Create Vector Mask}
  }
  Store the \VL columns as sparse\;
  \caption{Vectorized \SPARS algorithm.\label{algo.spars}}
 \end{algorithm}
 \vspace{-0.4cm}

 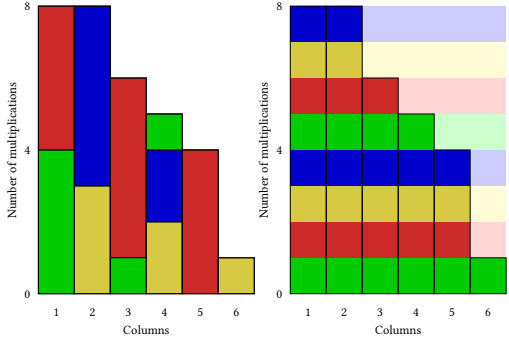
\begin{figure}

  \resizebox{0.8\linewidth}{!}{
  \begin{tikzpicture}
  \foreach \x in {1,...,6}
 {
   \node at (\x - 0.5, -0.5) {\x};
   \node at (\x + 6.5, -0.5) {\x};
  }
  \node at (-0.3, 0) {0};
  \node at (-0.3, 4) {4};
  \node at (-0.3, 8) {8};
  \node at (6.7, 0) {0};
  \node at (6.7, 4) {4};
  \node at (6.7, 8) {8};
  \node [rotate=90] at (-0.7, 4)  {Number of multiplications};
  \node [rotate=90] at (6.3, 4)  {Number of multiplications};
  \node at (3,-1) {Columns};
  \node at (10,-1) {Columns};
  \draw[fill=green!80!black] (0,4) rectangle (1,0);
  \draw[fill=red!80!black] (0,8) rectangle (1,4);
  \draw[fill=yellow!80!black] (1,3) rectangle (2,0);
  \draw[fill=blue!80!black] (1,8) rectangle (2,3);
  \draw[fill=red!80!black] (2,6) rectangle (3,1);
  \draw[fill=green!80!black] (2,1) rectangle (3,0);
  \draw[fill=yellow!80!black] (3,2) rectangle (4,0);
  \draw[fill=blue!80!black] (3,4) rectangle (4,2);
  \draw[fill=green!80!black] (3,5) rectangle (4,4);
  \draw[fill=red!80!black] (4,4) rectangle (5,0);
   \draw[fill=yellow!80!black] (5,1) rectangle (6,0);

  \fill[green!80!black] (7,1) rectangle (13,0);
  \fill[red!80!black] (7,2) rectangle (12,1);
  \fill[red!20!white] (12,2) rectangle (13,1);
  \fill[yellow!80!black] (7,3) rectangle (12,2);
  \fill[yellow!20!white] (12,3) rectangle (13,2);
  \fill[blue!80!black] (7,4) rectangle (12,3);
  \fill[blue!20!white] (12,4) rectangle (13,3);
  \fill[green!80!black] (7,5) rectangle (11,4);
  \fill[green!20!white] (11,5) rectangle (13,4);
  \fill[red!80!black] (7,6) rectangle (10,5);
  \fill[red!20!white] (10,6) rectangle (13,5);
  \fill[yellow!80!black] (7,7) rectangle (9,6);
  \fill[yellow!20!white] (9,7) rectangle (13,6);
  \fill[blue!80!black] (7,8) rectangle (9,7);
  \fill[blue!20!white] (9,8) rectangle (13,7);

  \draw (0,8) rectangle (1,0);
  \draw (7,8) rectangle (8,0);
  \draw (1,8) rectangle (2,0);
  \draw (8,8) rectangle (9,0);
  \draw (2,6) rectangle (3,0);
  \draw (9,6) rectangle (10,0);
  \draw (3,5) rectangle (4,0);
  \draw (10,5) rectangle (11,0);
  \draw (4,4) rectangle (5,0);
  \draw (11,4) rectangle (12,0);
  \draw (5,1) rectangle (6,0);
  \draw (12,1) rectangle (13,0);

  \end{tikzpicture}
  }
  \caption{Visual comparison of \SPA (left) and \SPARS (right).
  \label{fig.comparison_spa_spars}}
 \end{figure}

 Figure~\ref{fig.comparison_spa_spars} shows a graphical representation of both the \SPA and \SPARS algorithms when computing $C=A\times B$.
 The x-axis represents columns of the $C$ matrix and the y-axis the number of multiplications required to compute the non-zero elements of each $C$ column. 
 Colored blocks represent computations performed by the same vector instruction.
 \SPA uses vector instructions to compute each $C$ column vertically by loading columns of $A$ and multiplying their non-zero elements by the corresponding $b_{ij}$ coefficient, while \SPARS progresses horizontally in the $C$ matrix by processing a single multiplication
 per column at each step. 
 The vector length of \SPARS is defined by the number of columns that it simultaneously processes. 
 For example, in the case of Figure~\ref{fig.comparison_spa_spars}, \SPARS handles 6 columns concurrently, so \VL = 6.
 The potential of \SPARS for exploiting the length of long vector instructions is determined by the total colored area of Figure~\ref{fig.comparison_spa_spars}.
 The plain area (left and bottom) represents useful work,
 while the semitransparent area (top and right) corresponds to areas of the $C$ matrix with sparsity patterns
 that force \SPARS to use masks and leave a significant portion of the vector instruction length unused.

 Using a small block size reduces the length of vector instructions triggered by \SPARS and increases its total number of instructions.
 However, small block sizes have the benefit of reducing
 the number of masked elements within vector instructions.
 Conversely, increasing the block size introduces more masked elements per instruction, but it reduces the number of instructions and increases \VL.
 For example, if we divide the matrix of Figure~\ref{fig.comparison_spa_spars} in two blocks of 3 columns, \VL would experience a reduction from 6 to 3 and
 the number
 of instructions of \SPARS would increase from 8 to 13, with 8 instructions for the first block and 5 for the second block.
 In addition, we would avoid the top-right colored square of size $3\times3$, which would reduce the number of masked elements per instruction. 
 In general, \SPARS performs well when it processes large blocks that do not require the extensive use of masks.
 Adjacent C columns displaying very different numbers of non-zero coefficients may severely impact \SPARS performance since these differences force \SPARS to use masks.
We propose two optimizations to boost the performance of \SPARS: {\it Sorting}, which sorts matrix columns depending on their computational load to homogenize as much as possible consecutive columns; and a {\it Blocking Algorithm} to enable the use of the largest possible block sizes that do not contain large differences between their columns.
 

 \sloppy {\bf Sorting:} As a pre-processing phase, we sort the columns of $B$ in decreasing order in terms of computational load.
 \shepherd{The goal of this sorting phase is to decrease the load imbalance between vector lanes.}
 To determine the number of operations to compute each $B$ column $j$ we make the following argument: 
 each non-zero element $B_{kj}$ has to be multiplied by all the non-zero elements in column $k$ of $A$. If we denote by $Z^A_k$ the number
 of non-zero elements in column $k$ of $A$, we derive the number of operations $Op_{j}$ for column $j$:
 \[ Op_{j} = \sum\limits_{\substack{k\\B_{kj}\neq 0}} Z^A_k \]
 \shepherd{Then we sort all columns of $B$ by decreasing order of $Op_{j, 1\leq j\leq n_B}$, that is, we apply a permutation $P$ on matrix $B$.
 The result of the algorithm is now matrix $C \times P$ instead of $C$, which can be easily restored by keeping $P$ in memory.
 We do not actually reorder matrix $B$ but just the sequence of columns to be accessed.
 The non-zero elements of $B$ can be retrieved in
 the appropriate order by using indexed loads (Algorithm~\ref{algo.spars}, line 7). Matrix $A$ is left unchanged and requires no additional pre-processing.}
 
 {\bf Blocking Algorithm:} 
 Our algorithm expresses block sizes in terms of the vector length, that is, each block is processed using a vector length corresponding to the block size.
 We propose the following algorithm to dynamically determine block sizes when $j$ columns are already processed:
 \begin{enumerate}
  \item Start with \minb columns, from $j+1$ to $j_2=j+\minb$;
  \item While $Op_{j_2+1}$ is equal to $\max\limits_{j+1\leq j' \leq j_2} Op_{j'} = Op_{j+1}$, increment $j_2$;
  \item Stop the loop if $j_2$ reaches $j+\maxb$ or $n_B$.
 \end{enumerate}
 With this algorithm, columns $j+1$ to $j_2$ are processed within the same block, i.e. a block size of $j_2-j$.
 This algorithm introduces two new parameters: \minb is the minimum block size and \maxb is the maximum block size. Our algorithm ensures that all columns in the same block will have exactly the same number of operations to perform,
 unless there are less than \minb columns satisfying this condition. The impact of the two parameters is discussed in Section~\ref{sec.expe}.

 {\bf Complexity analysis:} 
We can easily analyze the complexity of both \SPA and \SPARS by assuming
that every column of $A$ has exactly $Z^A$ non-zero elements.
 Column $j$ of $B$ has $Z^B_j$ non-zeros and we assume that $B$ columns are sorted, $i. e.$, $Z^B_{j} >= Z^B_{j+1}, \forall j=1,...,n_{B}-1$. $B$ has a total of $NNZ^B$ non-zeros. Finally we assume that \VL divides the number of columns $n_B$.
 With these assumptions it is possible to derive the total number of iterations required by each algorithm and the total number of matrix coefficients processed, which can be either zero or non-zero.
 The zero elements are guarded with masks, and total number of non-zero elements is the same for both algorithms, since they both compute the same $C$ matrix.
 \SPA executes $NNZ^B$ 
 iterations of the main loop and processes a total of $Z^A NNZ^B$ elements. 
 \SPARS has a total 
 iteration count of $\sum\limits_{k=0}^{n_B/\VL-1} Z^A Z^B_{1+k\VL}$, which is bounded by $Z^A (\frac{NNZ^B}{\VL} + Z^B_1-Z^B_{n_B})$, and
 processes $\sum\limits_{k=0}^{n_B/\VL-1} \VL Z^A Z^B_{1+k\VL}$ elements, which is bounded by $ Z^A (NNZ^B + \VL( Z^B_1 - Z^B_{n_B}) )$.
 In the simplified case where all columns of $B$ have the same number of non-zeros, that is $Z^B_1 = Z^B_{n_B}$, both algorithms process the same number of elements.
In this scenario, the difference between \SPA and \SPARS in terms of iteration count comes from factor $\frac{Z^A}{\VL}$.
When $Z^A$ is small, \SPARS requires less iterations than \SPA, that is, \SPARS is better suited for very sparse matrices featuring a low number of non-zeros per column.
Otherwise, when $Z^A$ large, \SPA requires less iterations to compute the SpGEMM kernel than \SPARS, that is, \SPA is better suited for matrices that have a relatively large number of non-zeros per column.
Section~\ref{sec.expe.synthetic} empirically validates these theoretical observations.


\vspace{-0.2cm}
\subsection{HASH algorithm}
 \label{sec.algo.HASH}
 This section proposes \HASH, an algorithm to compute SpGEMM on vector processors that relies on hash functions to store intermediate values of the $C$ output matrix.
In contrast, \SPA and \SPARS rely on a dense structure to accumulate these intermediate values.
\HASH defines its hash table in terms of two arrays of size $H$: \texttt{HASH\_values} and \texttt{HASH\_indices}.
The former contains the values of non-zero elements for a given column, while the latter stores the corresponding row indices.
Instead of directly storing the intermediate products at the position given by row index $i$,
\HASH stores the value and the index at the position given by the hash function: $h(i) = ( i * c ) \text{ mod } H$, where $c$ is a constant.
However, collisions can occur if $h(i_1) = h(i_2)$ and $i_1 \neq i_2$. 
In this case, \HASH uses
linear probing to solve the collision, $i.e.$, it stores value and index in the next available cells of the hash table, going back to the beginning if needed.
As collisions create write dependencies and they can only be solved by sequentially accessing the hash table, vectorizing \HASH 
requires each B column to have its own hash table and to accumulate the intermediate product at each step.
The pseudo-code for \HASH is very similar to Algorithm~\ref{algo.spars}, with the accesses to the 3 SPA arrays replaced by hash function computations,
collision handling, and accesses to the hash table.
$B$ columns are sorted by decreasing load $Op_j$ and block sizes are determined using the same algorithm as in Section~\ref{sec.algo.SPARS} as well.
The motivation for these two optimizations is the same as the \SPARS algorithm.
The novelty of \HASH with respect to previous work on hash functions for vector SpGEMM~\cite{spgemm_vector} relies on two aspects:
i) \HASH uses the same blocking and sorting algorithms as \SPARS; ii) \HASH dynamically adapts the size of its hash table to optimize data transfers, which is possible thanks to the sorting pre-process. 

{\bf Hash table size adaptation:}
Previous works using hash functions in the context of SpGEMM computations set the hash table size at the beginning of the execution to a power of two~\cite{hash_example2, spgemm_vector}, so table size is strictly
larger than the maximum number of intermediate products in one column.
In the case of vector architectures,
setting the same maximum value throughout the whole SpGEMM execution can lead to severe performance degradation.
For this reason, we dynamically adapt the hash table size as we go through the different columns by using the maximum number of intermediate products in one column, $Op_j$, which we determine during preprocessing time.
%
%
We start with an initial size $H=2^k$, where $k$ is defined by the inequalities $2^{k-1} \leq \max\limits_j  Op_j < 2^k$. The $j$ indices correspond to all the columns of the first $B$ block.
For the next block of columns,it can happen that $\max\limits_j Op_j < 2^{k-l}$, for $j$ an index column of this block and $l$
an integer greater or equal than 1. When it happens, $H$ can be set to $2^{k-l}$, reducing the range of memory accesses. 
This process continues during the whole execution of \HASH. 

{\bf Complexity analysis:}
The complexity analysis of \HASH is similar to that of \SPARS, as intermediate products are generated in the same way.
However, there are several differences between both algorithms. 
The first difference is that \HASH requires computing hash function values while \SPARS directly accesses SPA arrays, which adds a small overhead.
The second difference comes from hash function collisions, which take place while \HASH runs.
A single collision among the \VL intermediate products that are computed in parallel at each step is enough to introduce some idle time to all the \VL computations.
If the probability of having a collision for one element is $p_c$, the probability of having
a collision in \HASH is $1-(1-p_c)^\VL$, which is close to $\VL \times p_c$ when $p_c << 1$. 
Finally, the third and major difference is the size of the \HASH table compared to the size of \SPARS dense vectors.
While the memory footprint of the \SPARS dense vector has an order of magnitude of $O(m^A \cdot \VL)$, \HASH reduces it to $O( \max\limits_j Op_j \cdot \VL)$ for very sparse matrices.

 \vspace{-0.2cm}
 \subsection{Hybrid algorithms}
 \label{sec.hybrid}
 
This section proposes the \SPARSH{t} and \HASHH{t} hybrid algorithms, which combine the benefits of \SPA with \SPARS and \HASH, respectively.  
As we state in Section~\ref{sec.algo.SPARS}, and results in Section~\ref{sec.expe.synthetic} empirically confirm, \SPARS performs better than \SPA for sparse matrices featuring a very small number of non-zeros per column, while \SPA delivers better performance than \SPARS when the number of non-zeros per column is relatively large. 
The same general conclusions hold for \HASH compared to \SPA.
\SPARSH{t} and \HASHH{t} require the same sorting pre-process as \SPARS and \HASH.
Using parameter $t$, \SPARSH{t} (or \HASHH{t}) dynamically switches between \SPA and \SPARS (or \HASH). 
Since columns $j$ are sorted by decreasing load ($Op_{j}$), \SPARSH{t} and \HASHH{t} process the first columns using \SPA as long as $Op_{j} >= t$.
Then, when $Op_{j} < t$, \SPARSH{t} (or \HASHH{t}) switches to \SPARS (or \HASH) to process the last columns by blocks.
The block size is determined by the blocking algorithm described in Section~\ref{sec.algo.SPARS}.
The two limit values for $t$ are $t=0$ and $t=\infty$. When $t=0$, \SPARSH{0} and \HASHH{0} process all columns using \SPA so \SPARSH{0}, \HASHH{0} and \SPA represent the same algorithm,
except that the columns are initially sorted in the case of \SPARSH{0} and \HASHH{0}. 
When $t=\infty$, all columns are processed in blocks, so \SPARSH{\infty} and \SPARS represent the same algorithm, while \HASHH{\infty} is the same as \HASH.

 \section{ESC}
 
 \label{sec.esc}
 In this Section, we briefly describe another commonly used algorithm in the litterature called \ESC~\cite{esc_gpu_dalton,winter_ESC}. This algorithm does not rely on the Gustavson method.
\ESC stands for \textit{Expand-Sort-Compress} and is split into three phases.
The first phase, \textit{Expand}, takes as input a group of $C$ columns and generates all the intermediate products from $A$ and $B$ needed to compute those $C$ columns in an array called \texttt{esc\_val}.
Along with the value of the product, the row and column indices of each one is also saved into two arrays \texttt{id\_row} and \texttt{id\_col}.
The second phase, \textit{Sort}, sorts \texttt{esc\_val} first by using \texttt{id\_row} as keys, then using \texttt{id\_col} as keys. At the end of this phase, the values with the same position in $C$ are contiguous and ordered column by column.
The last phase, \textit{Compress}, accumulates all the values that have the same pair of indices to create the non-zero elements of $C$. They can then be stored in CSC format.

\paragraph{Implementation.}
We implement \ESC in the following manner. In the \textit{Expand} phase, we generate $k$ triplets (\texttt{id\_row},\texttt{id\_col},\texttt{esc\_val}) from $n$ columns of B, as \SPA proceeds. One vector instruction
is used for each non-zero element of $B$ and the vector length corresponds to the number of non-zeros in the corresponding column of $A$.
The \textit{Sort} phase is implemented with a radix sort~\cite{radix_sort_vpu}, which consists in several rounds of sorting using a vectorized bucket sort.
Finally, the \textit{Compress} phase is done using strided memory instructions. Virtual processor $i$, with $i=0,\dots,\VL-1$, compresses triplets $\frac{k}{\VL}i$ to $\frac{k}{\VL}(i+1)-1$.
The possible compression of triplets handled by two different virtual processors is done through a sequential loop of a size \VL at the end. The resulting non-zero elements are stored sequentially into matrix $C$ to avoid relying
on slow prefix sum operations, as the final number of elements can be different for each virtual processor and column pointers for the CSC format are not easily set with a vectorized algorithm.

\paragraph{Optimizations.}
The execution of \ESC is highly determined by the execution time of the \textit{Sort} phase. It usually represents around 80-95\% of the total execution time of the algorithm.
Two main parameters determine the performance of the radix sort on long vector architectures: the number of elements to sort and the radix length.\\
Increasing the number of elements allows a better use of large vector lengths, as the algorithm only includes one sequential loop of size \VL.
Conversely, reducing the number of elements grants faster execution of indexed loads and stores, especially when the elements are re-ordered at the end of a round. Thus, there exists
an intermediate number of elements to achieve optimal performance. We thus determine how many columns are processed at each iteration of \ESC by grouping columns until the number of
intermediate products is above a given threshold, that we set to 10,000 in our experiments.\\
The radix length, denoted by $r$, also influences the performance of the radix sort. A larger radix length reduces the number of rounds to sort the elements. If the maximum
value of an index is $N$, the number of rounds is $\lceil \log_{2^r} N \rceil$. In the case of SpGEMM, the maximum value is either the number of columns or the number of rows of $C$.
However, in the vectorized algorithm, each virtual processor holds $2^r$ buckets, meaning that the total size of the array used for bucket sort is $\VL\cdot 2^r$. Again, in order to reduce
the range of memory addresses during indexed loads and stores, a low value of $r$ is preferable, even more so when \VL is large.
We experimentally determined that 5 or 6 is the best radix length to use a large vector length, thus we set it to $r=5$ unless $r=6$ decreases the number of sorting rounds.

%
%
%

%
 \vspace{-0.15cm}
 \section{Experimental evaluation}
 \label{sec.expe}
 This section provides an evaluation of our algorithms.
We describe our experimental setup in Section~\ref{sec.expe.setup}. 
We evaluate the performance of \SPA, \SPARS and \HASH using synthetic matrices in Section~\ref{sec.expe.synthetic}.
We perform an extensive evaluation of \SPA, \SPARS, \HASH, \SPARSH{t}, \HASHH{t} and \ESC.
the SuiteSparse Matrix Collection~\cite{suitesparseMC} in Section~\ref{sec.expe.results}.
We analyze the sensitivity of \SPARSH{t} and \HASHH{t} to parameters $t, \minb,$ and $\maxb$ in Section~\ref{sec.expe.sensitivity}.
Source code and evaluation files can be found at~\cite{zenodo}.


\subsection{Experimental Setup}
\label{sec.expe.setup}

{\bf Hardware Platform:} We run our experimental campaign on a vector processor architecture composed of a core and a Vector Processing Unit (VPU)~\cite{vitruvius}.
The core is a 6-stage 
single-issue RISC-V core.
It supports the execution of vector instructions on the VPU.
Vector instructions are decoded by the core and sent to the VPU for execution.
Scalar instructions run on the core.
The VPU has 8 vector lanes and supports a maximum \VL of 256 Double-Precision (DP) elements.
There is one Fused Multiply-Add (FMA) unit per lane capable of delivering 2 DP Floating-Point (FP) operations per cycle, which means that the whole VPU delivers up to 16 DP FP operations/cycle.
We use a FPGA to implement this vector processor architecture.
The FPGA is a VCU128 board running at 50MHz and implements the scalar core and the VPU accelerator. 
It has 4GB DDR4 DRAM and 1MB of L2 cache.

{\bf Sparse Matrices:} Our experimental campaign considers two sets of matrices. 
Section~\ref{sec.expe.synthetic} describes results considering a set of synthetic matrices where we specify the number of non-zero elements per column and we use a uniformly distributed random variable to distribute them across the column.
Experiments involving synthetic matrices make it possible to easily analyze the sensitivity of our algorithms to the sparsity of input matrices. 
Section~\ref{sec.expe.results} provides an extensive evaluation campaign of \SPA, \SPARS, \HASH, \SPARSH{t}, \HASHH{t} and \ESC considering a heterogeneous set of 40 sparse matrices obtained from the SuiteSparse Matrix Collection~\cite{suitesparseMC}.
Both Sections ~\ref{sec.expe.synthetic} and ~\ref{sec.expe.results} contain further details on the matrices and the algorithms being used.

{\bf Execution and Measurement Details} We compute $C=A\times B$ for the case of $A=B$.
Our performance results report the total execution time of the algorithm.
We do not account for pre-execution steps like memory allocation in our measurements, since they are orthogonal to the performance of the parallel algorithms we consider. 

\subsection{Performance on Synthetic Matrices}
\label{sec.expe.synthetic}

This section evaluates the performance of \SPA, \SPARS and \HASH using synthetic matrices.
These matrices make it possible to clearly display the sensitivity of these algorithms with respect to some matrix input parameters.
We generate synthetic matrices using two parameters: $n$, the number of rows and columns, and $Z$, the number of non-zero elements per column. 
A uniformly distributed random variable determines the specific positions of the $Z$ elements within each column.
Since the number of non-zero elements is constant in every column, the number of multiplications to perform per column is also constant and equal to $Z^2$. 
Thus, sorting columns is not required in these experiments.
Also, parameter \minb is not relevant either since the size of each block always reaches the maximum value $\maxb$.

\begin{figure}
 \includegraphics[width=0.9\linewidth]{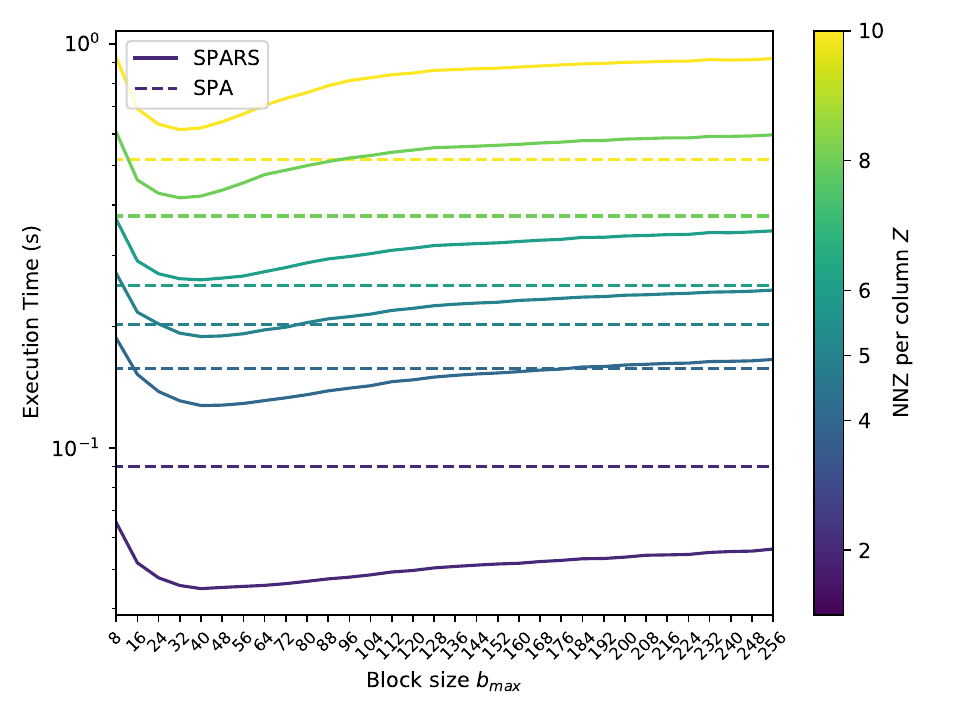}
 \vspace{-0.5cm}
 \caption{Performance of \SPA and \SPARS on synthetic matrices, for different $Z$ and $\maxb$.\label{fig.comparison_SPA_SPARS}}
 \vspace{-0.5cm}
\end{figure}

\begin{figure}
 \includegraphics[width=0.9\linewidth]{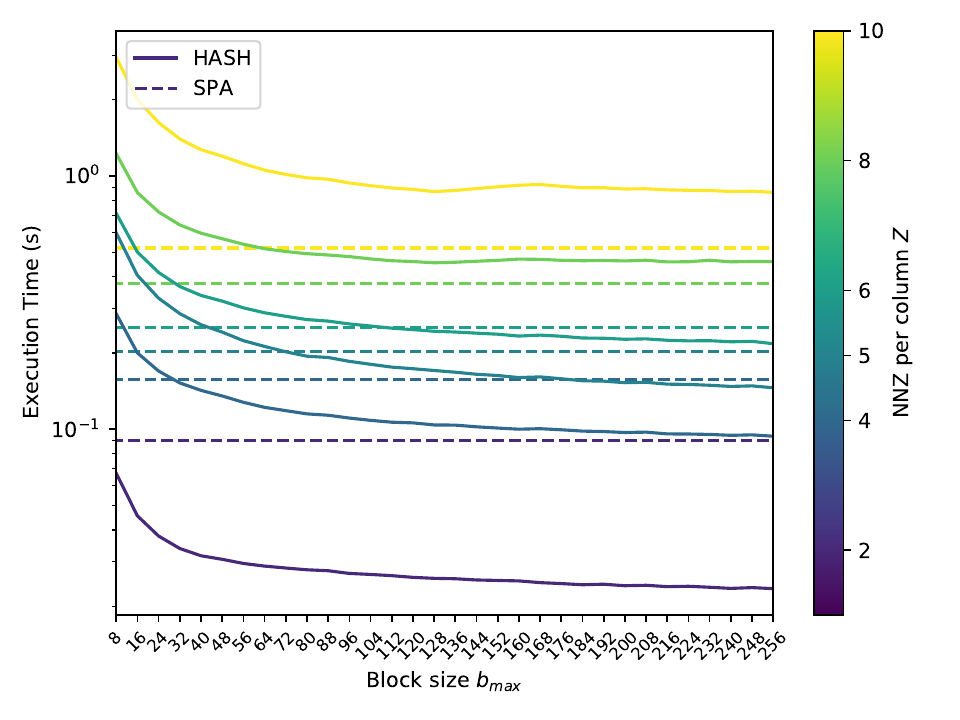}
 \vspace{-0.5cm}
 \caption{Performance of \SPA and \HASH on synthetic matrices, for different $Z$ and $\maxb$.\label{fig.comparison_SPA_HASH}}
\end{figure}

Figure~\ref{fig.comparison_SPA_SPARS} compares the performance of \SPA and \SPARS and Figure~\ref{fig.comparison_SPA_HASH} compares the performance of \SPA and \HASH.
The x-axis represents the value of the \maxb parameter. We consider \maxb values from 8 to 256. 
The matrix size $n$ is set to 2560, so that there are at least 10 blocks in all scenarios.
The y-axis represents execution time.
Solid lines represent the execution time of either \SPARS or \HASH while dotted lines represent \SPA.
We consider the following values for the number of non-zero elements per column, $Z$: $2,4,5,6,8$ and $10$. 
We represent these different values using different colors to show the performance of \SPARS (or \HASH) and \SPA. 
The legend in the right hand side of the plot shows the color we use to represent results considering different values of $Z$.



Results in Figure~\ref{fig.comparison_SPA_SPARS} indicate that \SPARS reaches its optimal performance at $\maxb=40$ and worsens for larger $\maxb$ values.
This effect is driven by the performance of indexed load and store instructions. 
Since large $\maxb$ values produce vector instructions with large $\VL$, setting up a large $\maxb$ value forces indexed loads and stores vector instructions to trigger a large number of memory requests,
over a large range of addresses, in a reduced time interval, which overwhelms the memory subsystem and induces performance degradation for large values of $\maxb$.
On the contrary, results in Figure~\ref{fig.comparison_SPA_HASH} show that \HASH is not affected by this performance issue and is faster than \SPARS for the largest $\maxb$.
This confirms that reducing the size of data structures containing partial accumulations of $C$ values is critical as it reduces the range of addresses that indexed loads and stores accesses.
Importantly, large $\maxb$ values allow \HASH to operate with large $\VL$, which delivers significant performance benefits.
Although the number of hash table collisions increases linearly with $\VL$, as Section~\ref{sec.algo.HASH} indicates, the performance of \HASH improves with large \VL. 
Results concerning \SPA always produce an horizonal line since parameter \maxb does not impact \SPA performance.

Results in Figures~\ref{fig.comparison_SPA_SPARS} and~\ref{fig.comparison_SPA_HASH} also show that for large values of $Z$, 
\SPA performs better than \SPARS and \HASH, while for low values of $Z$, \SPARS and \HASH are faster than \SPA.
The complexity analysis described in Section~\ref{sec.algo.SPARS} indicate this same trend.
In particular, when we set $\maxb=40$ to maximize the performance of \SPARS, then \SPARS is faster than \SPA if and only if $Z \leq 5$ for sparse matrices with a constant number of non-zero coefficients per columns.
Similarly, when we set $\maxb=256$ to maximize the performance of \HASH, then \HASH is faster than \SPA if and only if $Z \leq 6$.
For the general case this particular condition does not strictly hold since the distribution of non-zero elements among matrix columns strongly impacts the performance of the three algorithms.
Finally, we can see that \HASH is really efficient for very low values of $Z$, as it is able to capitalize on a very small hash table to achieve quick indexed loads and stores.
When $Z$ increases, \SPARS becomes faster than \HASH, though worse than \SPA, as index computation and collisions increase the overhead of the algorithm compared to direct indexing in the dense vectors.

\subsection{Evaluation on real matrices}
\label{sec.expe.results}

This section evaluates the performance of the \SPA, \SPARS, \HASH, \SPARSH{t}, \HASHH{t} and \ESC algorithms considering a set of 40 square matrices that belong to the SuiteSparse Matrix Collection~\cite{suitesparseMC}.
These matrices were selected to (i) fit in the available memory of our setup; and (ii) represent different sparsity patterns.
The matrices exhibit different average ratios of non-zero elements by column, which is an important aspect that determines \SPA, \SPARS and \HASH performance.
They also exhibit more or less variance in the distribution of elements over the columns, which also highlights the performance of the hybrid algorithms compared to \SPARS and \HASH.
Table~\ref{table.matrices} lists the set of 40 matrices.
The second and third columns display the number of rows and columns of each matrix (column $Size$) and the total number of non-zero elements per matrix (column $\#NNZ$), respectively.
The fourth, fifth, and sixth columns display the minimum, maximum, and average number of non-zero coefficients per column, respectively, while the seventh column displays the variance of the distribution of non-zero coefficients per column of each matrix.
Columns 8 to 11 display the same statistics but in terms of number of multiplications per column.

We consider the following ten algorithms:
\begin{compactitem}
 \item \EXPSPA, which is \SPARSH{0}.
 \item \EXPSPARSVAR, $i.e.$, \SPARSH{\infty} with a variable block size between $\minb=16$ and $\maxb=64$.
 \item \EXPSPARS, $i.e.$, \SPARSH{\infty} with a block size of $40$, $i.e.$, $\minb=\maxb=40$.
 \item \EXPSPARSHVAR, $i.e.$ \SPARSH{40}, $\minb=16$, $\maxb=64$.
 \item \EXPSPARSH, $i.e.$ \SPARSH{40}, $\minb=\maxb=40$.
 \item \EXPHASHVAR, $i.e.$ \HASHH{\infty}, $\minb=32$, $\maxb=256$.
 \item \EXPHASH, $i.e.$ \HASHH{\infty}, $\minb=\maxb=256$.
 \item \EXPHASHHVAR, $i.e.$ \HASHH{40}, $\minb=32$, $\maxb=256$.
 \item \EXPHASHH, $i.e.$ \HASHH{40}, $\minb=\maxb=256$.
 \item \EXPESC, as described in Section~\ref{sec.esc}.
\end{compactitem}



\begin{table*}
\resizebox{\linewidth}{!}{
\begin{tabular}{|l|cc|cccc|cccc|cccccccccc|}
\hline
\multirow{2}{*}{Name} & \multirow{2}{*}{Size} & \multirow{2}{*}{\#NNZ} & \multicolumn{4}{c|}{\#NNZ per column} & \multicolumn{4}{c|}{\#Multiplications per column} & \multirow{2}{*}{\EXPSPA (s)} & \multirow{2}{*}{\makecell{\textsc{Spars}\\16/64}} & \multirow{2}{*}{\makecell{\textsc{Spars}\\40/40}} & \multirow{2}{*}{\makecell{\textsc{H-Spa}\\16/64}} & \multirow{2}{*}{\makecell{\textsc{H-Spa}\\40/40}} & \multirow{2}{*}{\makecell{\textsc{Hash}\\32/256}} & \multirow{2}{*}{\makecell{\textsc{Hash}\\256/256}} & \multirow{2}{*}{\makecell{\textsc{H-Hash}\\32/256}} & \multirow{2}{*}{\makecell{\textsc{H-Hash}\\256/256}} & \multirow{2}{*}{\EXPESC} \\
\cline{4-11}
& & & Min & Max & Avg. & Var. & Min & Max & Avg.$\blacktriangle$ & Var. & & & & & & & & & & \\
\hline
\rowcolor{SPA} \poli & 4008 & 8188 & 1 & 15 & 2.04 & 0.46 & 1 & 38 & 3.92 & 5.83 & 1.50e-1 & 2.10$\times$ & 2.22$\times$ & 2.10$\times$ & 2.21$\times$ & \cellcolor{best}4.21$\times$ & 3.83$\times$ & 4.20$\times$ & 3.83$\times$ & 0.95$\times$ \\
\rowcolor{SPARS} \SPI & 2028 & 5007 & 0 & 8 & 2.47 & 0.30 & 0 & 25 & 6.39 & 1.50 & 8.69e-2 & 2.05$\times$ & 2.05$\times$ & 2.05$\times$ & 2.04$\times$ & \cellcolor{best}3.63$\times$ & 3.30$\times$ & 3.61$\times$ & 3.28$\times$ & 0.70$\times$ \\
\rowcolor{SPA} \koho & 4470 & 12731 & 0 & 51 & 2.85 & 10.20 & 0 & 221 & 11.88 & 238.58 & 2.32e-1 & 1.17$\times$ & 1.21$\times$ & 1.19$\times$ & 1.26$\times$ & 1.22$\times$ & 1.27$\times$ & 1.37$\times$ & \cellcolor{best}1.69$\times$ & 0.54$\times$ \\
\rowcolor{SPA} \hamr & 5952 & 22162 & 2 & 8 & 3.72 & 3.42 & 4 & 40 & 14.07 & 82.28 & 3.78e-1 & 1.29$\times$ & 1.42$\times$ & 1.29$\times$ & 1.42$\times$ & 2.26$\times$ & \cellcolor{sbest}2.31$\times$ & 2.25$\times$ & \cellcolor{best}2.32$\times$ & 0.59$\times$ \\
\rowcolor{SPA} \bp & 822 & 3276 & 1 & 20 & 3.99 & 10.43 & 1 & 107 & 14.18 & 272.39 & 4.97e-2 & 1.33$\times$ & 1.46$\times$ & 1.41$\times$ & \cellcolor{best}1.49$\times$ & 1.26$\times$ & 1.05$\times$ & 1.43$\times$ & 1.43$\times$ & 0.54$\times$ \\
\rowcolor{SPARS} \barth & 6019 & 23492 & 2 & 10 & 3.90 & 0.68 & 4 & 51 & 14.91 & 22.04 & 3.79e-1 & 1.36$\times$ & 1.48$\times$ & 1.36$\times$ & 1.48$\times$ & 2.27$\times$ & 2.29$\times$ & 2.28$\times$ & \cellcolor{best}2.33$\times$ & 0.57$\times$ \\
\rowcolor{SPARS} \oscil & 430 & 1544 & 1 & 13 & 3.59 & 2.33 & 1 & 60 & 15.00 & 65.90 & 2.43e-2 & 1.33$\times$ & 1.45$\times$ & 1.35$\times$ & \cellcolor{best}1.51$\times$ & 1.23$\times$ & 1.13$\times$ & 1.32$\times$ & 1.42$\times$ & 0.50$\times$ \\
\rowcolor{SPA} \rw & 5151 & 20199 & 1 & 4 & 3.92 & 0.11 & 2 & 16 & 15.49 & 3.1480 & 3.09e-1 & 1.32$\times$ & 1.40$\times$ & 1.32$\times$ & 1.40$\times$ & \cellcolor{sbest}2.20$\times$ & \cellcolor{best}2.21$\times$ & 2.19$\times$ & \cellcolor{best}2.21$\times$ & 0.53$\times$ \\
\rowcolor{SPA} \olm & 1000 & 3996 & 3 & 4 & 4.00 & 0.00 & 10 & 16 & 15.97 & 0.15 & 5.39e-2 & 1.55$\times$ & 1.48$\times$ & 1.55$\times$ & 1.48$\times$ & 2.15$\times$ & \cellcolor{best}2.18$\times$ & 2.12$\times$ & 2.16$\times$ & 0.51$\times$ \\
\rowcolor{SPA} \tub & 1000 & 3996 & 3 & 4 & 4.00 & 0.00 & 10 & 16 & 15.97 & 0.15 & 5.80e-2 & 1.68$\times$ & 1.60$\times$ & 1.68$\times$ & 1.60$\times$ & 2.29$\times$ & \cellcolor{best}2.33$\times$ & 2.28$\times$ & \cellcolor{sbest}2.32$\times$ & 0.56$\times$ \\
\rowcolor{SPARS} \bcsp & 1723 & 6511 & 2 & 15 & 3.78 & 3.02 & 5 & 80 & 17.30 & 102.80 & 1.10e-1 & 1.30$\times$ & 1.38$\times$ & 1.30$\times$ & 1.37$\times$ & 1.39$\times$ & 1.57$\times$ & 1.42$\times$ & \cellcolor{best}1.77$\times$ & 0.48$\times$ \\
\rowcolor{SPARS} \saylrt & 1000 & 3750 & 1 & 7 & 3.75 & 4.06 & 1 & 42 & 18.13 & 166.59 & 6.00e-2 & 1.25$\times$ & 1.38$\times$ & 1.26$\times$ & 1.36$\times$ & 1.66$\times$ & \cellcolor{best}2.03$\times$ & 1.63$\times$ & 1.92$\times$ & 0.53$\times$ \\
\rowcolor{SPARS} \shermanf & 1104 & 3786 & 1 & 7 & 3.43 & 6.40 & 1 & 47 & 18.16 & 332.27 & 5.77e-2 & 1.17$\times$ & 1.23$\times$ & 1.17$\times$ & 1.20$\times$ & 1.33$\times$ & \cellcolor{best}1.53$\times$ & 1.30$\times$ & 1.42$\times$ & 0.35$\times$ \\
\rowcolor{SPA} \qh & 1484 & 6110 & 2 & 13 & 4.12 & 2.56 & 5 & 68 & 19.51 & 94.54 & 9.71e-2 & 1.28$\times$ & 1.34$\times$ & 1.28$\times$ & 1.33$\times$ & 1.38$\times$ & 1.49$\times$ & 1.40$\times$ & \cellcolor{best}1.67$\times$ & 0.43$\times$ \\
\rowcolor{SPA} \shyy & 4720 & 20042 & 1 & 6 & 4.25 & 1.63 & 2 & 36 & 19.62 & 129.92 & 3.12e-1 & 1.26$\times$ & 1.38$\times$ & 1.26$\times$ & 1.38$\times$ & 2.16$\times$ & \cellcolor{best}2.23$\times$ & 2.16$\times$ & \cellcolor{best}2.23$\times$ & 0.48$\times$ \\
\rowcolor{SPA} \rajat & 7602 & 32653 & 1 & 52 & 4.29 & 1.26 & 3 & 303 & 19.71 & 51.70 & 5.15e-1 & 1.19$\times$ & 1.27$\times$ & 1.22$\times$ & 1.33$\times$ & 1.98$\times$ & 1.40$\times$ & 2.16$\times$ & \cellcolor{best}2.18$\times$ & 0.48$\times$ \\
\rowcolor{SPA} \young & 841 & 4089 & 3 & 5 & 4.74 & 0.21 & 11 & 25 & 22.51 & 11.03 & 5.85e-2 & 1.40$\times$ & 1.38$\times$ & 1.40$\times$ & 1.38$\times$ & 1.99$\times$ & \cellcolor{best}2.12$\times$ & 2.00$\times$ & \cellcolor{best}2.12$\times$ & 0.49$\times$ \\
\rowcolor{SPARS} \shermant & 5005 & 20033 & 1 & 7 & 4.00 & 7.09 & 1 & 49 & 23.11 & 411.19 & 3.36e-1 & 1.00$\times$ & 1.10$\times$ & 1.09$\times$ & 1.12$\times$ & 1.64$\times$ & \cellcolor{best}1.83$\times$ & 1.34$\times$ & 1.40$\times$ & 0.42$\times$ \\
\rowcolor{SPA} \dw & 2048 & 10114 & 3 & 8 & 4.94 & 0.26 & 11 & 49 & 24.54 & 17.05 & 1.52e-1 & 1.26$\times$ & 1.23$\times$ & 1.25$\times$ & 1.22$\times$ & 1.79$\times$ & \cellcolor{best}1.84$\times$ & 1.82$\times$ & \cellcolor{best}1.84$\times$ & 0.41$\times$ \\
\rowcolor{SPA} \rdb & 1250 & 7300 & 4 & 6 & 5.84 & 0.15 & 18 & 36 & 34.25 & 14.17 & 1.07e-1 & 1.21$\times$ & 1.17$\times$ & 1.21$\times$ & 1.17$\times$ & \cellcolor{best}1.64$\times$ & \cellcolor{sbest}1.63$\times$ & \cellcolor{sbest}1.63$\times$ & \cellcolor{sbest}1.63$\times$ & 0.33$\times$ \\
\rowcolor{SPA} \tols & 663 & 1712 & 1 & 22 & 3.25 & 25.97 & 1 & 471 & 38.00 & 13361.58 & 7.30e-2 & 0.92$\times$ & 0.79$\times$ & 1.36$\times$ & 1.36$\times$ & 0.70$\times$ & 0.35$\times$ & \cellcolor{best}1.52$\times$ & \cellcolor{best}1.52$\times$ & 0.25$\times$ \\
\rowcolor{SPA} \fpga & 1220 & 5852 & 1 & 36 & 4.80 & 20.44 & 7 & 164 & 38.12 & 427.76 & 1.09e-1 & 0.95$\times$ & 1.00$\times$ & 1.03$\times$ & 1.06$\times$ & 0.90$\times$ & 0.84$\times$ & 1.05$\times$ & \cellcolor{best}1.15$\times$ & 0.32$\times$ \\
\rowcolor{SPARS} \watt & 1856 & 11360 & 2 & 7 & 6.12 & 1.67 & 6 & 49 & 39.37 & 125.89 & 1.72e-1 & 1.08$\times$ & 1.08$\times$ & 1.05$\times$ & 1.05$\times$ & 1.36$\times$ & \cellcolor{best}1.39$\times$ & 1.14$\times$ & 1.13$\times$ & 0.35$\times$ \\
\rowcolor{SPARS} \saylrf & 3564 & 22316 & 3 & 7 & 6.26 & 0.56 & 13 & 49 & 39.76 & 52.96 & 3.55e-1 & 0.93$\times$ & 1.02$\times$ & 0.98$\times$ & 1.02$\times$ & 1.48$\times$ & \cellcolor{best}1.61$\times$ & 1.16$\times$ & 1.20$\times$ & 0.37$\times$ \\
\rowcolor{SPA} \ors & 2205 & 14133 & 4 & 7 & 6.41 & 0.41 & 19 & 49 & 41.49 & 49.78 &  2.06e-1 & 1.04$\times$ & 1.04$\times$ & 1.00$\times$ & 1.00$\times$ & 1.55$\times$ & \cellcolor{best}1.59$\times$ & 1.19$\times$ & 1.20$\times$ & 0.33$\times$ \\
\rowcolor{SPA} \wang & 2903 & 19093 & 4 & 7 & 6.58 & 0.37 & 19 & 49 & 43.62 & 46.98 & 2.93e-1 & 1.01$\times$ & 1.07$\times$ & 1.01$\times$ & 1.03$\times$ & 1.52$\times$ & \cellcolor{best}1.56$\times$ & 1.11$\times$ & 1.12$\times$ & 0.35$\times$ \\
\rowcolor{SPA} \gemat & 4929 & 33044 & 1 & 28 & 6.70 & 11.56 & 1 & 206 & 45.27 & 735.35 & 6.12e-1 & 0.85$\times$ & 0.93$\times$ & 0.99$\times$ & 1.02$\times$ & 0.79$\times$ & 0.95$\times$ & 1.06$\times$ & \cellcolor{best}1.10$\times$ & 0.37$\times$ \\
\rowcolor{SPARS} \lshp & 3466 & 23896 & 4 & 7 & 6.89 & 0.20 & 21 & 49 & 47.74 & 20.56 & 3.44e-1 & 0.94$\times$ & 1.01$\times$ & 0.98$\times$ & 0.98$\times$ & 1.46$\times$ & \cellcolor{best}1.48$\times$ & 1.00$\times$ & 1.00$\times$ & 0.31$\times$ \\
\rowcolor{SPA} \gresley & 4908 & 30482 & 2 & 34 & 6.21 & 9.39 & 8 & 324 & 48.25 & 1065.07 & 5.03e-1 & 0.79$\times$ & 0.86$\times$ & 0.99$\times$ & 1.02$\times$ & 1.04$\times$ & 1.00$\times$ & 1.17$\times$ & \cellcolor{best}1.20$\times$ & 0.32$\times$ \\
\rowcolor{SPA} \lns & 3937 & 25407 & 1 & 13 & 6.45 & 10.39 & 1 & 113 & 48.44 & 866.46 & 4.00e-1 & 0.82$\times$ & 0.89$\times$ & 0.99$\times$ & 1.01$\times$ & \cellcolor{sbest}1.22$\times$ & \cellcolor{best}1.23$\times$ & 1.06$\times$ & 1.07$\times$ & 0.32$\times$ \\
\rowcolor{SPA} \pores & 1224 & 9613 & 2 & 30 & 7.85 & 29.53 & 10 & 298 & 63.62 & 2199.05 & 1.50e-1 & 0.78$\times$ & 0.89$\times$ & 1.01$\times$ & 1.01$\times$ & 0.77$\times$ & 0.59$\times$ & \cellcolor{best}1.03$\times$ & 1.01$\times$ & 0.29$\times$ \\
\rowcolor{SPA} \cheby & 6435 & 51480 & 3 & 9 & 8.99 & 0.02 & 15 & 65 & 64.92 & 2.12 & 5.23e-1 & 0.94$\times$ & 1.01$\times$ & 1.01$\times$ & 1.01$\times$ & \cellcolor{best}1.36$\times$ & \cellcolor{best}1.36$\times$ & 1.00$\times$ & 1.00$\times$ & 0.31$\times$ \\
\rowcolor{SPA} \str & 363 & 3068 & 1 & 26 & 8.45 & 84.35 & 1 & 449 & 70.61 & 12314.86 & 4.93e-2 & 0.83$\times$ & 0.91$\times$ & 1.02$\times$ & \cellcolor{best}1.05$\times$ & 0.65$\times$ & 0.25$\times$ & 0.99$\times$ & 0.93$\times$ & 0.32$\times$ \\
\rowcolor{SPA} \dwt & 2680 & 25026 & 4 & 19 & 9.34 & 3.44 & 27 & 228 & 90.65 & 623.75 & \cellcolor{sbest}4.01e-1 & 0.70$\times$ & 0.76$\times$ & \cellcolor{sbest}1.00$\times$ & \cellcolor{best}1.01$\times$ & 0.77$\times$ & 0.91$\times$ & \cellcolor{sbest}1.00$\times$ & 0.99$\times$ & 0.26$\times$ \\
\rowcolor{SPA} \cage & 3534 & 41594 & 3 & 23 & 11.77 & 14.08 & 15 & 474 & 152.60 & 7046.5951 & \cellcolor{best}8.00e-1 & 0.65$\times$ & 0.73$\times$ & \cellcolor{best}1.00$\times$ & \cellcolor{best}1.00$\times$ & 0.57$\times$ & 0.59$\times$ & \cellcolor{best}1.00$\times$ & \cellcolor{best}1.00$\times$ & 0.25$\times$ \\
\rowcolor{SPA} \nasa & 1824 & 39208 & 6 & 42 & 21.50 & 49.58 & 65 & 1197 & 511.64 & 59420.46 & \cellcolor{best}8.14e-1 & 0.41$\times$ & 0.47$\times$ & \cellcolor{best}1.00$\times$ & \cellcolor{sbest}0.99$\times$ & 0.36$\times$ & 0.31$\times$ & \cellcolor{sbest}0.99$\times$ & \cellcolor{sbest}0.99$\times$ & 0.16$\times$ \\
\rowcolor{SPA} \exdd & 839 & 22460 & 7 & 62 & 26.77 & 190.67 & 176 & 2270 & 907.22 & 220428.89 & 5.50e-1 & 0.33$\times$ & 0.41$\times$ & 1.00$\times$ & \cellcolor{sbest}1.01$\times$ & 0.29$\times$ & 0.20$\times$ & \cellcolor{best}1.02$\times$ & 1.00$\times$ & 0.17$\times$ \\
\rowcolor{SPA} \adder & 1813 & 11156 & 1 & 1332 & 6.15 & 1076.11 & 2 & 9439 & 1014.45 & 396265.13 & \cellcolor{best}2.25 & 0.61$\times$ & 0.64$\times$ & \cellcolor{best}1.00$\times$ & \cellcolor{best}1.00$\times$ & 0.34$\times$ & 0.18$\times$ & \cellcolor{best}1.00$\times$ & \cellcolor{best}1.00$\times$ & 0.20$\times$ \\
\rowcolor{SPA} \goodwin & 1965 & 56059 & 5 & 62 & 28.53 & 224.66 & 138 & 2359 & 1048.69 & 316412.44 & \cellcolor{best}1.47 & 0.31$\times$ & 0.38$\times$ & \cellcolor{best}1.00$\times$ & \cellcolor{best}1.00$\times$ & 0.27$\times$ & 0.24$\times$ & \cellcolor{best}1.00$\times$ & \cellcolor{sbest}0.99$\times$ & 0.14$\times$ \\
\rowcolor{SPA} \iprob & 3001 & 9000 & 2 & 3000 & 3.00 & 2994.00 & 3002 & 6000 & 3003.00 & 2994.00 & \cellcolor{best}10.33 & 0.77$\times$ & 0.72$\times$ & \cellcolor{best}1.00$\times$ & \cellcolor{best}1.00$\times$ & 0.34$\times$ & 0.31$\times$ & \cellcolor{best}1.00$\times$ & \cellcolor{sbest}0.99$\times$ & 0.18$\times$ \\
\hline
\multicolumn{11}{|r|}{Average speed-up} & 1.00$\times$ & 1.079$\times$ & 1.131$\times$ & 1.204$\times$ & 1.235$\times$ & 1.436$\times$ & 1.413$\times$ & 1.535$\times$ & \cellcolor{best}1.569$\times$ & 0.399$\times$ \\
\hline
\multirow{2}{*}{Name}& \multirow{2}{*}{Size} & \multirow{2}{*}{\#NNZ} & Min & Max & Avg. & Var. & Min & Max & Avg.$\blacktriangle$ & Var. & \multirow{2}{*}{\EXPSPA (s)} & \multirow{2}{*}{\makecell{\textsc{Spars}\\16/64}} & \multirow{2}{*}{\makecell{\textsc{Spars}\\40/40}} &\multirow{2}{*}{\makecell{\textsc{H-Spa}\\16/64}} & \multirow{2}{*}{\makecell{\textsc{H-Spa}\\40/40}} & \multirow{2}{*}{\makecell{\textsc{Hash}\\32/256}} & \multirow{2}{*}{\makecell{\textsc{Hash}\\256/256}} & \multirow{2}{*}{\makecell{\textsc{H-Hash}\\32/256}} & \multirow{2}{*}{\makecell{\textsc{H-Hash}\\256/256}} & \multirow{2}{*}{\EXPESC} \\
\cline{4-11}
 &  & & \multicolumn{4}{c|}{\#NNZ per column} & \multicolumn{4}{c|}{\#Multiplications per column} & & & & & & & & & & \\
\hline
\end{tabular}
}
\caption{Statistics and performance results on 40 sparse matrices.\label{table.matrices}}
\vspace{-0.3cm}
\end{table*}

Table~\ref{table.matrices} reports the execution time of \EXPSPA in terms of seconds in its 12th column, and the speed-ups obtained by the other algorithms with respect to \EXPSPA in its 9 rightmost columns.
Highlighted cells indicate the best algorithm(s) per matrix.
Matrices appear in Table~\ref{table.matrices} sorted in increasing average number of multiplications per column.
Results are consistent with the experiments with synthetic matrices of Section~\ref{sec.expe.synthetic} and the complexity analysis of Sections~\ref{sec.algo.SPARS} and ~\ref{sec.algo.HASH}: \SPARS-based and \HASH-based algorithms (\EXPSPARS, \EXPSPARSVAR, \EXPHASH, \EXPHASHVAR) deliver good performance for matrices featuring a small
number of non-zeros per column.
Conversely, \SPA is faster than \EXPSPARS, \EXPSPARSVAR, \EXPHASH and \EXPHASHVAR when dealing with denser matrices.
\EXPSPARS has an average speed-up of 1.13$\times$ for the whole set of matrices with respect to \EXPSPA. 
It reaches a
1.38$\times$ speed-up when considering the 22 most sparse matrices of the set, which feature on average less than 6 non-zeros per column and less than 39 multiplications per column.
\EXPSPARSVAR obtains 1.08$\times$ average speed-up for the 40 matrices and reaches 1.34$\times$ for the 22 most sparse ones.
\EXPHASH and \EXPHASHVAR have an average speed-up of 1.41$\times$ and 1.44$\times$, respectively, for the 40 matrices. 
For the 22 most sparse ones, their average speed-up is 1.85$\times$ and 1.88$\times$ respectively.
\HASH-based algorithms are usually faster than \SPARS-based for very sparse matrices. For matrices where \EXPSPA is the fastest algorithm, this is the opposite, which confirms the remarks made in Section~\ref{sec.expe.synthetic}.

\EXPSPARS generally delivers better performance than \EXPSPARSVAR. 
\EXPSPARS forces the block size to be 40 to target the best performance, which constitutes a better strategy than forming blocks with similar load between $\minb=16$ and $\maxb=64$, as \EXPSPARSVAR does.
There are two reasons explaining why \EXPSPARS is faster than \EXPSPARSVAR: (i) since many columns share the same value $Op_{j}$, \EXPSPARSVAR usually forms blocks of size 64 ($\maxb$ value), which is not the optimal block size as Section~\ref{sec.expe.synthetic} indicates, and 
(ii) although \EXPSPARS leaves some portions of the vector processor idle for some time, it compensates this idle time by processing more columns than \EXPSPARSVAR with less instructions.
In average, \EXPHASHVAR has a slightly larger speed-up than \EXPHASH, but \EXPHASH is more often the best algorithm among these two. 
Forcing a block size of 256 allows to compute as many intermediate products as possible in one instruction,
and it is the value which gives the best performance when there is little variation in the number of non-zero elements per column, as we show in Section~\ref{sec.expe.synthetic}.

Compared to \EXPHASH, the state-of-the-art vectorized hash algorithm for SpGEMM~\cite{spgemm_vector} takes on average 52\% more time when using a block size of 256. 
The lowest speed-up that \EXPHASH obtains with respect to this previous approach is 1.01$\times$ for matrices
\olm, \rdb and \cheby, which have very small variances.
The highest is 4.49$\times$ for \rajat, where columns have between 3 and 303 intermediate products.

Additionally, the preprocessing time is often less than 40\% of the execution time of \HASH algorithms, and even smaller than 5\% for the 6 last matrices. Only 5 matrices report a relative preprocessing time
greater than 40\%: \poli, \SPI, \hamr, \barth and \rw. For these 5 matrices \EXPHASH and \EXPHASHVAR have good speed-ups.

\begin{figure*}
 \subfigure[Varying $t$]{\includegraphics[width=0.28\linewidth]{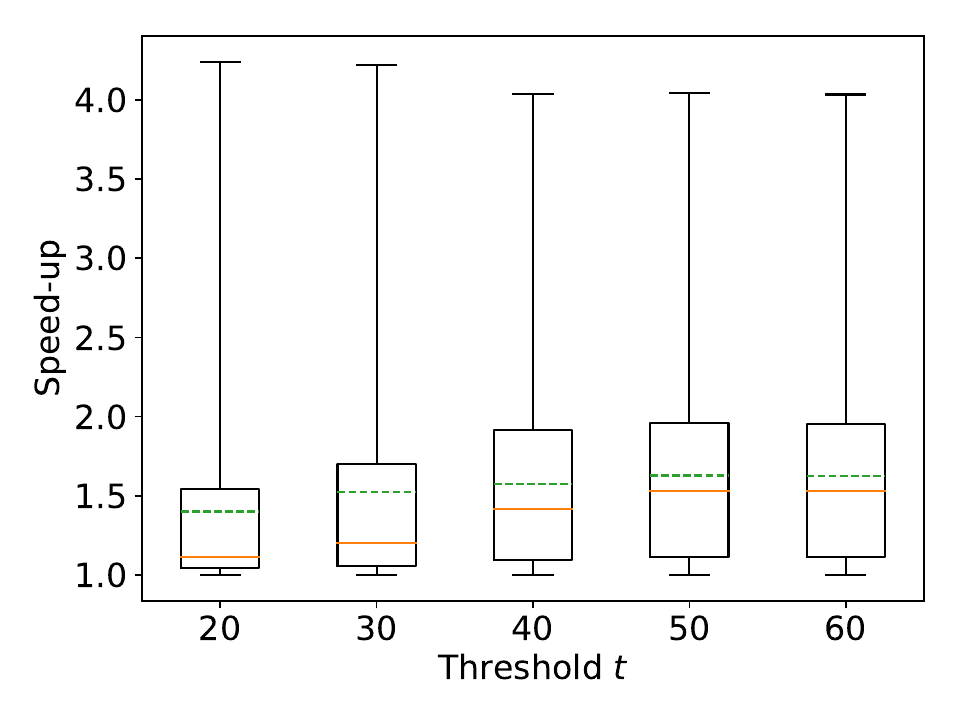}\label{fig.sens_hash.t}}\hfill
 \subfigure[Varying \minb]{\includegraphics[width=0.28\linewidth]{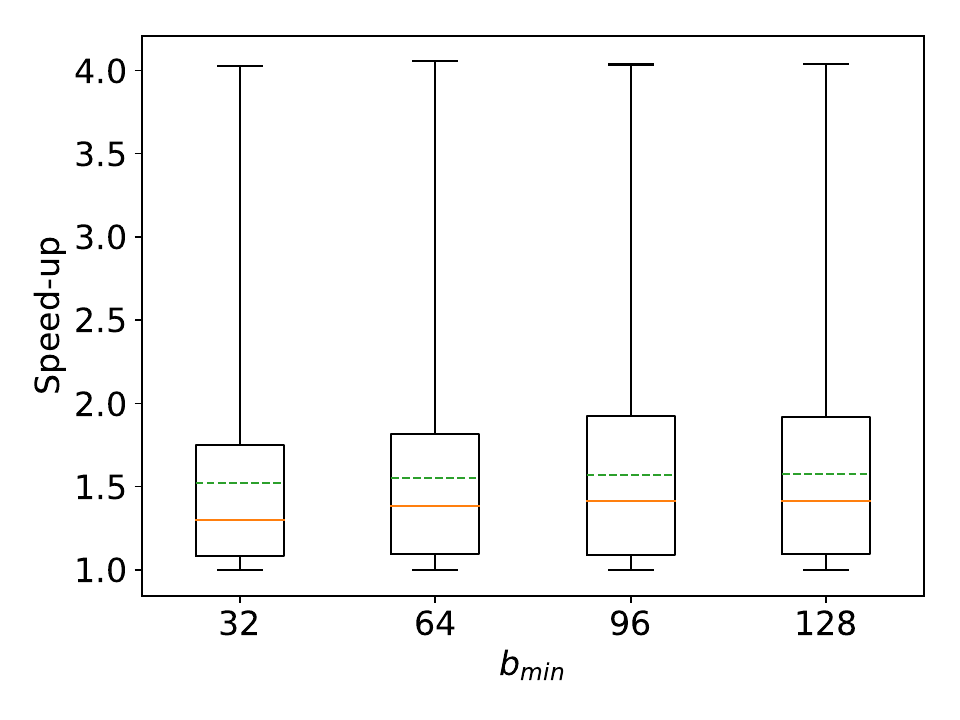}\label{fig.sens_hash.minb}}\hfill
 \subfigure[Varying \maxb]{\includegraphics[width=0.28\linewidth]{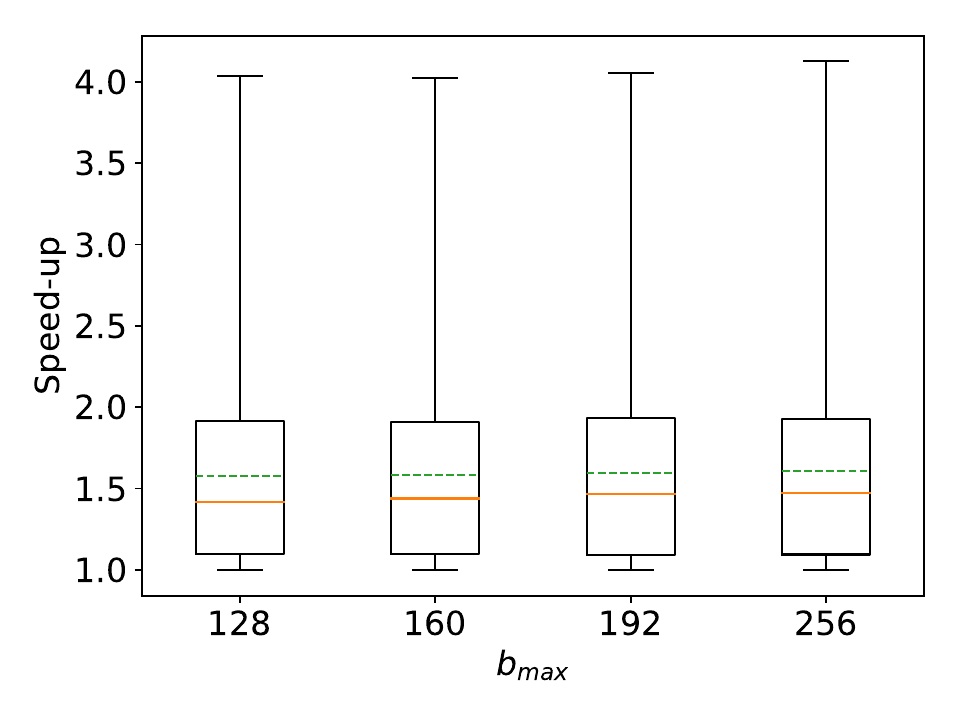}\label{fig.sens_hash.maxb}}
 \caption{Speed-ups of \HASHH{t} with respect to \SPA considering different values of the parameters $t$ (left), \minb (middle) and \maxb (right).\label{fig.sensitivity_hash}}
 \vspace{-0.2cm}
\end{figure*}

\EXPSPARSH obtains a 1.24$\times$ average speed-up with respect to \SPA over the whole set of 40 matrices.
For the 22 most sparse matrices, \EXPSPARSH delivers a 1.42$\times$ average speed-up.
Table~\ref{table.matrices} shows that \EXPSPARSH is able to largely outperform \SPA, \EXPSPARS, and \EXPSPARSVAR.
For matrix blocks with a large computational load per column, \EXPSPARSH is able to dynamically adapt its strategy and use \SPA, which allows it to outperform \EXPHASH and \EXPHASHVAR.
For matrix blocks with low computational load, it is able to switch to \SPARS, which remains slower than \HASH-based algorithms. 
\EXPSPARSH reaches particularly large speed-ups with respect to \SPA, \EXPSPARS, and \EXPSPARSVAR in matrices featuring a large variance across columns in terms of the number of non-zero values. 
Neither \SPA nor \SPARS-based approaches achieve good performance in these matrices. 
For example, \EXPSPARSH reaches a large speed-up of 1.36$\times$ for \tols with respect to \SPA.
It also largely outperforms \EXPSPARS and \EXPSPARSVAR, as well as \EXPHASH and \EXPHASHVAR for this matrix. 
Table~\ref{table.matrices} indicates that \EXPSPARSHVAR delivers similar performance as \EXPSPARSH. \EXPSPARSHVAR reaches an average 1.20$\times$ speed-up with respect to \EXPSPA over the whole set of 40 matrices and it delivers its largest speed-ups with respect \EXPSPA for the 22 most sparse matrices, as \EXPSPARSH does.
The performance of \EXPSPARSHVAR is slightly worse than \EXPSPARSH for the same reason as \EXPSPARSVAR is worse than \EXPSPARS, $i.e.$, setting $\minb$ and $\maxb$ to 40 creates better matrix blocks than setting $\minb$ to 16 and $\maxb$ to 64.

The same conclusions hold for \EXPHASHH and \EXPHASHHVAR, as they reach speed-ups with respect to \EXPSPA of 1.57$\times$ and 1.54$\times$ respectively. 
They are able to switch to \SPA for the densest matrices, which makes it possible to avoid the performance slowdowns that \EXPHASH and \EXPHASHVAR suffer for these matrices.
Their average speed-up is higher than \EXPSPARSH and \EXPSPARSHVAR because \HASH-based algorithms outperforms \SPARS-based algorithms for very sparse matrices. 
In particular, \EXPHASHH is able to reach a speed-up of $1.99\times$ for the 22 most sparse matrices.
\EXPHASHH delivers slightly better performance than \EXPHASHHVAR because \EXPHASH is often faster than \EXPHASHVAR for the most sparse matrices.

\EXPESC is a slow algorithm for vector processors. Even though this algorithm does not suffer from load imbalance, sorting partial outputs is an expensive operation compared to accumulating values.
\EXPESC achieves an average speed-up of only 0.40$\times$ compared to \EXPSPA over the 40 matrices.

\vspace{-0.3cm}
\subsection{Sensitivity Analysis of \HASHH{t}}
\label{sec.expe.sensitivity}

This section provides a sensitivity analysis of the \HASHH{t} algorithm performance with respect to parameters $\minb$, $\maxb$, and $t$.
While Section~\ref{sec.expe.results} considers two pairs $(\minb,\maxb)$, $(32,256)$ and $(256,256)$, and a single threshold $t=40$ for \HASHH{t},
Figure~\ref{fig.sensitivity_hash} shows the distribution of the speed-ups achieved by \HASHH{t} over the 40 matrices in Table~\ref{table.matrices} with respect to \SPA for different parameter settings.
The orange solid line represents the median, the green dotted line represents the mean, the edges of the box represent the first and third quartiles, while the whiskers indicate the minimum and maximum speed-ups achieved.
We have similar results for the \SPARSH{t} but we omit them since they show a very similar sensitivity pattern to parameters $\minb$, $\maxb$, and $t$ as \HASHH{t}.

Figure~\ref{fig.sens_hash.t} shows the distribution of speed-ups for \HASHH{t} with $\maxb=\minb=128$ and $t=20,30,40,50,60$.
Increasing the threshold $t$ brings a small reduction of the largest speed-ups.
However, a small threshold does not benefit the majority of matrices, as with $t=20$, 24 matrices have speed-ups lower than $1.20\times$, compared
to only 12 with $t=50,60$.
The average speed-ups are 1.40$\times$ for $t=20$, 1.52$\times$ for $t=30$, 1.57$\times$ for $t=40$, 1.63$\times$ for $t=50$ and 1.62$\times$ for $t=60$.
The medians follow the same trend, and the interval $[40,60]$ is the best for the overall performance.

Figure~\ref{fig.sens_hash.minb} shows the distribution of speed-ups for \HASHH{t} with $\maxb=128$, $\minb=32,64,96,128$ and $t=40$.
The distributions are quite similar for all values of \minb.
The average speed-ups are 1.52$\times$ for $\minb=32$, 1.55$\times$ for $\minb=64$, 1.57$\times$ for $\minb=96$ and 1.58$\times$ for $\minb=128$.
In conclusion, an overall small increase in performance can be seen when \minb increases,
as a larger \VL benefits the \HASH algorithm.

Figure~\ref{fig.sens_hash.maxb} shows the distribution of speed-ups for \HASHH{t} with $\maxb=128,160,192,256$, $\minb=128$ and $t=40$.
The average speed-ups are 1.58$\times$ for $\maxb=128$, 1.58$\times$ for $\maxb=160$, 1.59$\times$ for $\maxb=192$ and 1.61$\times$ for $\maxb=256$.
Similarly to the study of \minb, increasing \maxb allows bigger blocks of columns to be processed at the same time. The impact is small since
block sizes between \minb and \maxb are only used when the number of multiplications
per column is exactly the same, thus increasing the minimal block size \minb has more impact than increasing \maxb.

 \vspace{-0.15cm}
 \section{Related work}
 \label{sec.related}
 {\bf Load imbalance}
Load imbalance is a frequently mentioned problem of the Gustavson method, especially on GPU.
Parger et al.~\cite{dense_gpu} use a low complexity pre-processing analysis of the matrices, linear in the number of non-zeros.
Depending on the result, a binning method can be used to reduce load imbalance of the algorithm on a GPU.
Liu et al.~\cite{38bins} compute the number of multiplications in each row and use it as an upper bound on the final
non-zero count of the row. They partition the rows into 38 bins depending on this upper bound, which
guides execution choices in terms of algorithm and allocation space. 
Nagasaka et al.~\cite{nagasaka_loadbalancing} map matrix rows to the different threads while enforcing that the total number of floating-point operations 
is equal for each one.
Our proposals \SPARS, \HASH and hybrid algorithms mitigate the load imbalance coming from differences between matrix columns by applying the sorting technique we describe in Section~\ref{sec.algo.SPARS}.

{\bf Alternatives to dense vectors}
Deveci et al.~\cite{hash_example2} use both hash maps and dense vectors as sparse accumulators. 
Nagasaka et al.~\cite{nagasaka_loadbalancing} present two alternatives based on hash maps. The first one is a classic algorithm~\cite{hash_example},
where column indices represent keys and linear probing is used in the case of collision. 
The second alternative
uses vector registers to increase the number of stored elements in each cell of the hash map. This reduces the number of collisions
at the cost of higher overhead.
\shepherd{\HASHH{t} incorporates three fundamentally novel aspects with respect to these previous approaches: i) it reduces load imbalance by sorting $B$ columns; ii) it dynamically adapts the hash table size; and iii) it dynamically switches between algorithms based on dense vector acummulators ($i.e.$ \SPA) and techniques based on hash maps ($i.e.$ \HASH).
Our experiments indicate that, without these innovations, the HASH-based algorithm delivers worse performance than \SPA on our long vector platform (0.95$\times$ degradation w.r.t. \SPA).
Remarkably, our algorithms ($e.g.$ \HASHH{t}) can be applied to multi-core CPU devices equipped with SIMD units~\cite{sodani16} in a similar way as previous work~\cite{nagasaka_loadbalancing}, and deliver performance improvements.
}

Another method for accumulating non-zeros in a column is the \emph{Expand-Sort-Compress (ESC)}~\cite{sort_bell}.
This method \textit{expands} the list of partial products with their corresponding row and column indices, \textit{sorts} them,
and finally \textit{compresses} all entries with the same row and column indices. This method is used 
in Cusp~\cite{cusp}.
Winter et al.~\cite{winter_ESC} use ESC. They use a two-level load balancing scheme to
distribute the load across GPU threads. Dalton et al.~\cite{esc_gpu_dalton} use
the same sorting method as us to reduce load imbalance in their fine-grain ESC implementation.
Kunchum et al.~\cite{kunchum_esc} propose a variant of ESC. The goal is to optimize the performance
for $A$ rows displaying a high number of non-zero compared to their corresponding $B$ rows.a
\shepherd{Section~\ref{sec.expe.results} evaluates the ESC algorithm and shows that it delivers poor performance in the context of long vector architectures.}


{\bf Non Gustavson-based approaches}
Lee et al.~\cite{block_reorganizer} use the outer-product formulation and design a 
block reorganizer algorithm to improve load balancing between GPU threads.
Niu et al.~\cite{spgemm_tiled} have recently proposed a tiled algorithm for SpGEMM to avoid
the most common problems of the Gustavson method on GPUs. They also have proposed a threshold to switch
between a dense and a sparse representation of each tile and designed a tiled algorithm for SpMV~\cite{spmv_tiled}.
Finally, Xie et al. presented a novel AI-based approach for SpGEMM~\cite{IA_spgemm}. Although the algorithms on which they rely
all use the Gustavson method, they investigate different matrix storage formats and algorithms and train a custom deep-learning network
which chooses the best solution for given input matrices and target architecture.
This paper uses a Gustavson-based approach to compute SpGEMM since it exposes a high degree of parallelism to the vector architecture.

{\bf Parallel Execution of SpMV and SpMM}
The parallel execution of other sparse linear algebra kernels like SpMV (Sparse Matrix-Vector multiplication) has been explored over the years.
The main difference between the SpMV operation $y=A \times x$ and SpGEMM is that the $x$ vector is dense.
Load imbalance undermines SpMV performance~\cite{spmv_mergepath}. Li et al.~\cite{spmv_analysis} provide a detailed
performance analysis of SpMV considering different matrix storage formats on a GPU.
More recently, approaches to efficiently run SpMV on long vector architectures have been proposed~\cite{spmv_vector}.

SpMM performs the multiplication between a sparse and a \textit{dense} matrix.
While it is possible to compute SpMM by running SpMV multiple times~\cite{spmm_spmv}, that is, by considering the dense matrix as several dense vectors, 
some optimizations based on coalescing data accesses over the whole matrix have been proposed~\cite{spmm_coalescing}. 
Koanantakool et al.~\cite{spmm_comm} provide lower bounds and optimal algorithms to efficiently run SpMDM on distributed memory scenarios.
Azad et al.~\cite{azad_spgemm}
present a similar analysis for SpGEMM, which has been extended to include memory constraints~\cite{spgemm_comm}.

{\bf Numerical Algorithms on Vector Processors}
Methods to efficiently run relevant numerical algorithms on vector architectures have been previously proposed.
For example, efficiently running the dense general matrix-matrix multiplication using vector instructions has been studied by Lim et al.~\cite{gemm_vector}. They use a blocking
version of the algorithm coupled with AVX-512 instructions and carefully chosen block sizes to execute
the micro-kernel with high performance. This work is extended by Kim et al.~\cite{gemm_vector_2}, which propose
an auto-tune mechanism that explores a reduced search space considering several parameters like
block size or prefetch distances, among others. Quicksort and Bitonic sort have also been
optimized for the AVX-512 ISA~\cite{quicksort_vector}, while stencil computations have been optimized for ARM SVE by Armejach et al.~\cite{stencil_vector}. Zhong et al.~\cite{bosilca_mpi_vector} investigate the use of vector intrinsics
in the context of MPI reductions, for both the AVX-512 and SVE ISAs.

Afanasyev et al. proposed Breadth-First Search~\cite{bfs_vector} and Connected Components~\cite{ConnectedComponents_vector} graph algorithms for the SX-Aurora TSUBASA vector architecture~\cite{sx-aurora}, which features registers able to store 256 double-precision values.
Li et al.~\cite{spgemm_vector} propose a SpGEMM algorithm based on \SPA targeting this same long vector architecture.
This paper proposes algorithmic innovations that deliver larger performance than \SPA, particularly for very sparse matrices.
 \section{Conclusion}
 \label{sec.conclusion}
\SPA-based approaches constitute the state-of-the-art to run SpGEMM on vector architectures, and deliver good performance for matrices featuring at least seven non-zero elements per column.
Since these approaches are not able to fully exploit the potential of long vector architectures for very sparse matrices, we present two algorithms able to efficiently manipulate very sparse matrices with a few non-zero elements on vector architectures, \SPARS and \HASH.
To successfully combine \SPA and \SPARS (or \HASH) and deliver good performance for a wide range of sparsity scenarios, we present \SPARSH{t} (or \HASHH{t}). It exploits a hybrid execution model that dynamically switches between \SPA and \SPARS (or \HASH).
In particular, \shepherd{\HASHH{t} is generally the best algorithm and} obtains a 1.57$\times$ average speed-up with respect to \SPA over a heterogeneous set of 40 sparse matrices.
For the 22 most sparse matrices, \HASHH{t} delivers a 1.99$\times$ average speedup over \SPA.
In the future we plan to propose methods to further accelerate the SpGEMM kernel on long vector architectures by exploiting data reuse at the vector register file.

 \section*{Acknowledgement}

This research has received funding from the European High Performance Computing Joint Undertaking (JU) under Framework Partnership Agreement No 800928 (European Processor Initiative) and Specific Grant Agreement No 101036168 (EPI SGA2). The JU receives support from the European Union’s Horizon 2020 research and innovation programme and from Croatia, France, Germany, Greece, Italy, Netherlands, Portugal, Spain, Sweden, and Switzerland. The EPI-SGA2 project, PCI2022-132935 is also co-funded by MCIN/AEI /10.13039/501100011033 and by the UE NextGenerationEU/PRTR. Marc Casas has been partially
supported by the Grant RYC-2017-23269 funded by MCIN/AEI/10.13039/501100011033 and by ESF Investing in your future. This research was supported by grant PID2019-107255GB-C21 funded by MCIN/AEI/ 10.13039/501100011033. Els autors agraeixen el suport del Departament de Recerca i Universitats de la Generalitat de Catalunya al Grup de Recerca "Performance understanding, analysis, and simulation/emulation of novel architectures" (Codi: 2021 SGR 00865).
 
\bibliographystyle{ACM-Reference-Format}
 \bibliography{biblio.bib}
 
\end{document}